%% file: arxiv_tech_report.tex
\documentclass[times,twocolumn,final]{elsarticle}

\usepackage{graphicx}
\usepackage{mathtools}
\usepackage[caption=false]{subfig}
\usepackage[all]{nowidow}
\usepackage{siunitx}
\usepackage{nth}

\usepackage{flushend}

\usepackage[utf8]{inputenc}
\usepackage{algorithm}
\usepackage{algpseudocode}
\usepackage{listings}
\usepackage{pgfplots}
\usepackage{amsmath}
\usetikzlibrary{patterns}
\pgfplotsset{width=10cm,compat=1.9}
\usepackage{balance}

\usepackage{multicol}

\pgfplotsset{compat=1.11,
    /pgfplots/ybar legend/.style={
    /pgfplots/legend image code/.code={%
       \draw[##1,/tikz/.cd,yshift=-0.25em]
        (0cm,0cm) rectangle (9pt,0.8em);},
   },
}

\usepackage{framed,multirow}

\usepackage{amssymb}
\usepackage{latexsym}

\usepackage{url}
\usepackage{xcolor}
\definecolor{newcolor}{rgb}{.8,.349,.1}

\usepackage{hyperref}

\usepackage[switch,pagewise]{lineno} 

\journal{arxiv}

\begin{document}


\begin{frontmatter}

\title{Compact Tetrahedralization-based Acceleration Structure for Ray Tracing}%

\author[1]{Aytek~Aman}
\ead{aytek.aman@cs.bilkent.edu.tr}
\author[1]{Serkan Demirci}
\ead{serkan.demirci@bilkent.edu.tr}
\author[1]{U\u{g}ur G\"{u}d\"{u}kbay\corref{cor1}}
\ead{gudukbay@cs.bilkent.edu.tr}
\cortext[cor1]{Corresponding author: 
  Tel.: +90-312-290-1386;  \\
  fax: +90-312-266-4047;}

\address[1]{Department of Computer Engineering, Bilkent University, 06800, Ankara, Turkey}


\begin{abstract}
We propose a compact and efficient tetrahedral mesh representation to improve the ray-tracing performance. We reorder tetrahedral mesh data using a space-filling curve to improve cache locality. Most importantly, we propose an efficient ray traversal algorithm. We provide details of common ray tracing operations on tetrahedral meshes and give the GPU implementation of our traversal method. We demonstrate our findings through a set of comprehensive experiments. Our method outperforms existing tetrahedral mesh-based traversal methods and yields comparable results to the traversal methods based on the state of the art acceleration structures such as \textit{k}-dimensional (\textit{k}-d) trees and Bounding Volume Hierarchies (BVHs). 
\end{abstract}


\end{frontmatter}


\section{Introduction}
\label{intro}

\textcolor{black}{The core operation in ray tracing is the ray-surface intersection calculations, which may contribute more than 95\% of the total computation time~\cite{Glassner:1989:Raytracing}. Hence, the computational cost of intersection calculations determines the run-time efficiency of the ray-tracing algorithm. To speed up this operation, the most common approach is to use spatial subdivision structures that partition the scene so that the triangles are enclosed in different volumes. During ray-traversal, ray-triangle intersection tests can be avoided if the enclosing volume for a triangle does not intersect with the ray. For partitioning the scene, regular grids, octrees, Bounding Volume Hierarchies (BVH), and \textit{k}-dimensional (\textit{k}-d) trees are commonly used. BVHs and \textit{k}-d trees are the most preferred space partitioning structures for ray tracing, thanks to the recent advancements in the construction and traversal methods.}

A more recent alternative to accelerate ray-surface intersection calculations is to use tetrahedralizations. A tetrahedral mesh is a three-dimensional (3-D) structure that partitions the 3-D space into tetrahedra. Constrained tetrahedralizations are a special case of tetrahedralizations that take the input geometry into account. In the resulting tetrahedral mesh, the components of the input geometry such as faces, line segments, and points are preserved. Similar to their 2-D counterparts, tetrahedralizations can be constructed in such a way that they exhibit Delaunay property; i.e., the tetrahedra are close to regular. There are three categories of constrained tetrahedralizations: \textit{Conforming Delaunay Tetrahedralization}, \textit{Constrained Delaunay Tetrahedralization}, and \textit{Quality Delaunay Tetrahedralization}~\cite{Lagae:2008:Constrained}.

Lagae and Dutr\'{e}~\cite{Lagae:2008:Constrained} use constrained tetrahedral meshes for rendering typical 3-D scenes. They tetrahedralize the space between objects in a constrained manner where the triangles in the scene geometry align with the triangles of the tetrahedral mesh. Then, they calculate ray-triangle intersections by traversing the tetrahedral mesh. Because a tetrahedral mesh is not a hierarchical structure, ray-surface intersections are mostly calculated by traversing a few tetrahedra. Besides, this approach has the advantages of providing a unified data structure for global illumination, handling deforming geometry if the topology (connectivity) of the mesh does not change, easily applying level-of-detail approaches, and ray tracing on the Graphics Processing Unit (GPU)~\cite{Lagae:2008:Constrained}.

Despite these advantages, the state-of-the-art traversal methods for tetrahedral meshes, such as Scalar Triple Product (ScTP), are still a magnitude or two slower than the k-d tree-based traversal, as Lagae and Dutr\'{e} state~\cite{Lagae:2008:Constrained}. We aim to improve the performance of the tetrahedral mesh-based traversal for ray-tracing as follows.

\begin{itemize}
    \item We propose a compact tetrahedral mesh representation to improve cache locality and to utilize memory alignment.
    \item We sort tetrahedral mesh data (tetrahedra and points) using a space-filling curve to improve cache locality.
    \item We propose an efficient tetrahedral mesh traversal algorithm using a modified basis that reduces the cost of point projection, which is frequently used during traversal.
    \item We utilize the GPU to speed up the ray-surface intersection calculations.
\end{itemize}

Additionally, we propose a simple technique to associate vertex attributes (normals, texture coordinates, and so on) with the tetrahedral mesh data.
Through experiments, we observe that our method performs better than the existing tetrahedral mesh-based traversal methods in terms of the computational cost. In certain scenes, especially the scenes with challenging geometry where there are long, extended triangles, we observe a better rendering performance than the \textit{k}-d tree and BVH implementations of the pbrt-v3~\cite{Matt:2016:Pbr}. Although this method cannot replace and improve upon the state-of-the-art accelerators (such as BVHs and \textit{k}-d trees) because of its disadvantages in its current form, its orthogonal strengths compared to the alternatives make it valuable and promising. This is especially important for aggregate structures where accelerators with different advantages can be combined to have the best of both worlds.

\section{Related Work} 
\label{sec:related_work}

\subsection{Acceleration Structures}

First proposed by Fujimoto et al.~\cite{Fujimoto:1988:Accelerated}, a regular grid partitions the 3-D scene into equally-sized boxes where each box keeps a list of triangles. During traversal, some well-known algorithms such as the three-dimensional digital differential analyzer (3D DDA) can be used to quickly determine the boxes that intersect with the ray. Although there are compact and robust acceleration structures such as the one proposed in~\cite{Lagae:2008:Grid}, one major disadvantage of the regular grid is its non-adaptive structure. The majority of the grid cells may not contain any triangles, while some grid cells may have a large number of triangles, which increases the average traversal cost.

One of the popular structures in the literature is BVHs. A BVH is a collection of hierarchical bounding volumes that enclose the objects in the scene. BVHs improve the ray tracing performance by culling the scene geometry using bounding volume intersection tests. Therefore, less triangle-ray intersection tests have to be performed compared to the brute force full scene traversal. Modern BVH construction techniques employ Surface Area Heuristic (SAH)~\cite{Goldsmith:1987:Hierarchies} to construct acceleration structures that perform well. The state of the art BVHs are constructed using a greedy top-down plane-sweeping algorithm proposed by Goldsmith et al~\cite{MacDonald:1990:Subdivision}, which is extended by Stitch et al.~\cite{Stich:2009:Splits} using spatial splits. Wodniok et al.~\cite{Wodniok:2016:Subtree} use recursive SAH values of temporarily constructed SAH-built BVHs to reduce ray traversal cost further.

The octree is another spatial indexing structure that is used to accelerate ray tracing~\cite{Glassner:1984:Subdivision}. It divides the space into eight subspaces in a recursive manner. During ray tracing, the octree is used to index the scene into subspaces and it is useful to determine the subspaces that intersect with the rays. After an octree is constructed, triangles that reside in these subspaces can be queried and the closest intersection with the rays can be found by performing a relatively small number of ray-surface intersection tests compared to the brute force approach.

Similar to the octree, the \textit{k}-d tree is also a space partitioning structure that divides the space into two sub-spaces at each level by alternating the split axis. To reduce the average ray traversal cost on a \textit{k}-d tree, these split planes are selected using the SAH, which is proposed by~\cite{MacDonald:1990:Subdivision}. SAH-based \textit{k}-d tree construction approaches are later improved by~\cite{Havran:2002:RayShooting}. Wald et al.~\cite{Wald:2006:FastKd} propose a SAH-based \textit{k}-d tree construction scheme with 
$O(N \log N)$ computational complexity. The \textit{k}-d trees constructed using the SAH are adaptive to the scene geometry. This means that if a ray is not in the proximity of any scene geometry, only a few tree nodes are traversed. This reduces the computation cost of ray tracing on scenes where primitives in the scene are not uniformly distributed, which is a common scenario for 3-D scenes.

In many ray tracing applications, rays share a common point such as rays that originate from the camera or rays that are cast to the light sources after ray-surface intersections. The structures that are discussed above do not exploit the characteristics of such rays in ray tracing. There exist better structures that take advantage of rays that share a common point in space and creates indices accordingly. Light Buffer~\cite{Haines:1986:LightBuffer} is an approach that partitions the scene according to one light source in the scene, which is then used for shadow testing. Hunt et al.~\cite{Hunt:2008:Adaptive} propose the perspective tree, which is similar to the Light Buffer that uses a 3-D grid in a perspective space considering the position of the light source or the camera as root. They later improve the perspective tree approach using an adaptive splitting scheme using SAH~\cite{Hunt:2008:RaySpecialized}.

\subsection{Tetrahedral Mesh Construction and Traversal}

Given an input geometry, a tetrahedral mesh can be constructed using well-known algorithms in computational geometry. TetGen~\cite{Si:2015:TetGen} is a commonly used tool to generate tetrahedral meshes. TetGen uses Bowyer-Watson~\cite{Bowyer:1981:Dirichlet, Watson:1981:Delaunay} and the incremental flip~\cite{Edelsbrunner:1992:Flipping} algorithms. Both methods have the worst-case complexity of $O(N^2)$. If points are uniformly distributed in space, the expected run-time complexity is $O(N \log N)$. To ensure numerical robustness, Shewchuk's robust geometric predicates~\cite{Shewchuk:1996:Predicates} are used.

There are tetrahedral mesh-based traversal methods used for accelerating ray and surface intersection calculations in raytracing three-dimensional scenes.
Lagae and Dutr\'{e}~\cite{Lagae:2008:Constrained} use ScTP to traverse the tetrahedral mesh. Their method requires the computation of three to six ScTP to determine the exit face. ScTP computation involves a cross product followed by a dot product on 3-D vectors. Maria et al.~\cite{Maria:2017:Traversal} propose a fast tetrahedral mesh traversal method, which uses an efficient exit face determination algorithm based on Pl\"{u}cker coordinates. 

Our method uses an efficient traversal method that works in 2-D, resulting in very few floating-point operations per tetrahedron compared to these alternatives. Our data structure is also compact and memory aligned. We also use a space-filling curve to further improve cache locality. Maria et al.~\cite{Maria:2017:Convex, Maria:2014:Topological} also propose a new acceleration structure for ray tracing, constrained convex space partition (CCSP), as an alternative to tetrahedral mesh-based acceleration structures. CCSP is more suitable for architectural environments because such a partitioning of a scene contains a smaller number of convex volumes, rather than a large number of tetrahedra.

\subsection{Raycasting for Direct Volume Rendering}
Direct volume rendering methods for rendering irregular grids, mostly represented as unstructured tetrahedral volumetric meshes, rely on raycasting and the composition of shades of samples along the rays throughout the volume to calculate pixel colors. For example, Silva et al.~\cite{Silva:1996:LazySweep,Silva:1997:LazySweep} use a sweeping plane first applied in the x-z plane, and then a sweeping line applied on the z-axis. They process these sweep lines further to render volumetric data stored as an irregular grid. Berk et al.~\cite{Berk:2003:Direct} focus on the usage of hybrid methods to utilize the strengths of image- and object-space methods. They rely on a next-cell operation for determining the next tetrahedron that the ray travels, as proposed by Koyamada et al.~\cite{Koyamada:1992:Fast}.

Garrity~\cite{Garrity:1990:Irregular} uses a simple traversal method where the ray is intersected with tetrahedra faces and the closest intersection gives the exit face for the tetrahedron. Koyamada~\cite{Koyamada:1992:Fast} uses two (on average) point-in-triangle tests in 2-D to determine the exit face. Riberio et al~\cite{Ribeiro:2007:Memory} use a more compact data structure for reduced memory usage during traversal. They also utilize ray coherence to reduce run-time memory usage. Later on, they~\cite{Maximo:2008:Memory} improved this method by providing a hardware implementation with additional arrangements of the data structure for reduced memory usage. Marmitt and Slusallek~\cite{Marmitt:2006:Traversal} use a method proposed by Platis and Theoharis~\cite{Platis:2003:Plucker}, which employs Pl\"{u}cker Coordinates of the ray and the tetrahedron edges to determine the exit face. They use the entry face information to reduce the number of tests to determine the exit face. They find the exit face using 2.67 ray-line orientation tests per tetrahedron on average.

We aim to provide a fast and compact acceleration structure to quickly find ray and surface intersections for rendering three-dimensional scenes composed of polygons (surface data). As opposed to direct volume visualization methods, our acceleration structure can handle queries for random rays scattered in different directions, given that their origin is already located (ray connectivity). Direct volume rendering techniques are geared towards rendering volumetric data from a certain camera position and orientation. Our tetrahedral ray traversal scheme could be adapted to direct volume rendering methods for better cache utilization and reduced computational cost. Besides, the compact tetrahedral mesh representation we propose could be utilized for direct volume rendering to reduce memory requirements of unstructured tetrahedral meshes.

\section{Tetrahedral Mesh Representation} 
\label{sec:compact_tet_mesh_rep}

We use a compact tetrahedral mesh representation for better cache utilization. We store tetrahedral mesh in two arrays as proposed by Lagae and Dutr\'{e}~\cite{Lagae:2008:Constrained}. The first array stores the point data and the second array stores the tetrahedron data. Figure~\ref{fig:basic_tet_mesh_representation} depicts the tetrahedron data representation for typical scenarios.

\begin{figure}[htbp]
\begin{center}
\begin{tabular}{|c|c|c|c|}
\hline 
  $V_0^i$ & $V_1^i$ & $V_2^i$ & $V_3^i$ \\ \hline
  $N_0^i$ & $N_1^i$ & $N_2^i$ & $N_3^i$ \\   
\hline
\end{tabular}
\end{center}
\caption{Typical tetrahedron representation in the memory. $V_j^i$ represents the index of the $j$'th vertex of the $i$'th tetrahedron. $N_j^i$ represents the neighboring tetrahedron index, which is across the vertex $V_j^i$. Each field is an integer and four bytes long. Thus, the full tetrahedron data occupies 32 bytes of memory.}
\label{fig:basic_tet_mesh_representation}
\end{figure}

Instead of using this representation, we propose three tetrahedron storage schemes that are more compact and better suited for efficient traversal: \textit{Tet32}, \textit{Tet20}, and \textit{Tet16}, which are 32, 20, and 16 bytes, respectively. We store a common field, exclusive-or sum (xor-sum), in all these structures, inspired by xor linked list structures for reducing the memory requirements of linked lists~\cite{Sinha:2005:DoublyLinkedList}. Mebarki uses a similar structure for compact 2-D triangulations~\cite{Mebarki:2018:Xor}. $\textit{VX}^i$ denotes the xor-sum of the vertex indices of the $i^{th}$ tetrahedron and $V_j^i$ denotes the index of the $j^{th}$ vertex of the $i^{th}$ tetrahedron. We compute the xor-sum as follows.

\[ VX^i = V_0^i \oplus V_1^i \oplus V_2^i \oplus V_3^i \]

\textit{Tet32} structure contains the first three vertex indices, xor-sum of all vertex indices, and four neighbor indices. Its memory layout is depicted in Figure~\ref{fig:tet32_representation}. 

\begin{figure}[htbp]
  \begin{center}
  \begin{tabular}{|c|c|c|c|}
  \hline 
    $V_0^i$ & $V_1^i$ & $V_2^i$ & $ \textit{VX}^i$ \\ \hline
    $N_0^i$ & $N_1^i$ & $N_2^i$ & $N_3^i$ \\   
  \hline
  \end{tabular}
  \end{center}
  \caption{\textit{Tet32} structure. Each field is an integer and four bytes long. Thus, the full tetrahedron data occupies 32 bytes of memory.}
  \label{fig:tet32_representation}
\end{figure}

With the \textit{Tet32} representation, we can use the $xor$ operation to quickly retrieve the index of the vertex that is not on a given face. We can get the index of the fourth point $V_3^i$ as follows. 

\[ V_3^i = V_0^i \oplus V_1^i \oplus V_2^i \oplus \textit{VX}^i. \]

\noindent This follows from the fact that xor operation is associative, commutative, and has the property $X \oplus X = 0$.

In \textit{Tet20}, we get rid of vertex indices and store only the xor-sum and the neighboring indices. We use the xor-sum field to get the index of the unshared vertex of the next tetrahedron during traversal. To do this, shared vertices between two tetrahedra must be known. This is guaranteed by \textit{ray connectivity}, meaning that the start and endpoints of rays are always connected in a typical ray-tracing scenario. However, we use a \textit{source tet}, a tetrahedron with complete index information, to initialize the indices at the beginning. We can choose this tetrahedron randomly. Starting from \textit{source tet}, it is possible to reconstruct the indices of the neighboring tetrahedra. It should be noted that we need to sort the neighbor indices in a tetrahedron using their corresponding vertex indices to find the neighbor for a given vertex index. Figure~\ref{fig:tet20_representation} shows the memory representation of the \textit{Tet20} structure.

\begin{figure}[htbp]
  \begin{center}
  \begin{tabular}{|c|c|c|c|c|}
  \hline 
    $\textit{VX}^i$ & $N_0^i$ & $N_1^i$ & $N_2^i$ & $N_3^i$ \\
  \hline
  \end{tabular}
  \end{center}
  \caption{\textit{Tet20} structure. Each field is an integer and four bytes long. Thus, the full tetrahedron data occupies 20 bytes of memory.}
  \label{fig:tet20_representation}
\end{figure}

In \textit{Tet16}, instead of storing four neighbor indices explicitly, we store three values that can be used to reconstruct neighbor indices, given that the previous (neighbor) tetrahedron index is known. We compute these three indices as follows.

\[ \textit{NX}_0^i = N_0^i \oplus N_3^i \]
\[ \textit{NX}_1^i = N_1^i \oplus N_3^i \]
\[ \textit{NX}_2^i = N_2^i \oplus N_3^i \]

Knowing the index of a neighbor tetrahedron and its order, we can reconstruct the rest of the neighbors easily. For example, If we have $N_2^i$, we retrieve $N_1^i$ as follows.

\[ N_2^i = N_2^i \oplus \textit{NX}_2^i \oplus \textit{NX}_1^i \]

\noindent The resulting \textit{Tet16} structure is given in Figure \ref{fig:tet16_representation}.

\begin{figure}[htbp]
  \begin{center}
  \begin{tabular}{|c|c|c|c|}
  \hline 
    $\textit{VX}^i$ & $\textit{NX}_0^i$ & $\textit{NX}_1^i$ & $\textit{NX}_2^i$ \\
  \hline
  \end{tabular}
  \end{center}
  \caption{\textit{Tet16} structure. Each field is an integer and four bytes long. Thus, the full tetrahedron data occupies 16 bytes of memory.}
  \label{fig:tet16_representation}
\end{figure}

If the corresponding face is a part of the scene geometry, neighbor index data points to a structure, called \textit{constrained face}. We use a single bitmask to identify such faces on the neighbor tetrahedron index field where the remaining 31 bits are used to reference either a neighboring tetrahedron or a constrained face depending on the value of the bitmask. Constrained face structure holds a reference to the actual triangle geometry and stores references to the neighboring two tetrahedra indices. These indices are used to recover and initialize the tetrahedron data when scattering rays are to be traced. It should be noted that multiple constrained faces can point to a single triangle when we allow triangles to be subdivided during tetrahedralization to enable high-quality tetrahedral meshes.

\section{Tetrahedron Traversal} 
\label{sec:fast_ray_traversal}

As the first step of tetrahedron traversal, we construct a 2-D basis $b=(\vec{u}, \vec{v})$ from the ray direction using the method described in~\cite{Duff:2017:Orthonormal}. Then, we define a new 2-D coordinate system $C$ with basis $b$ and origin $o$ where $r_o$ is the ray origin. We transform tetrahedron vertices to the coordinate system $C$ to obtain four points in 2-D. We determine the exit face in the initial tetrahedron using at most four points in triangle tests in 2-D. The query point is at the origin because the ray origin is the center of the new coordinate system $C$. Once we determine the exit face, we keep the 2-D coordinates and indices of the points of the exit face as $p_0, p_1, p_2$ and $\textit{idx}_0, \textit{idx}_1, \textit{idx}_2$, respectively. We also fetch the next tetrahedron index using the neighbor data.

After the initialization step, we start traversing the tetrahedral mesh. We first fetch the index of the fourth corner of the next tetrahedron (three of them are already known because two neighboring tetrahedra share three vertices) using the following expression where $XV^{next}$ denotes the xor sum of the next tetrahedron. 

\[ \textit{idx}_3 = \textit{idx}_0 \oplus \textit{idx}_1 \oplus \textit{idx}_2 \oplus \textit{XV}^{\textit{next}} \]

\begin{figure}[ht!]
\centering
\includegraphics[width=\columnwidth]{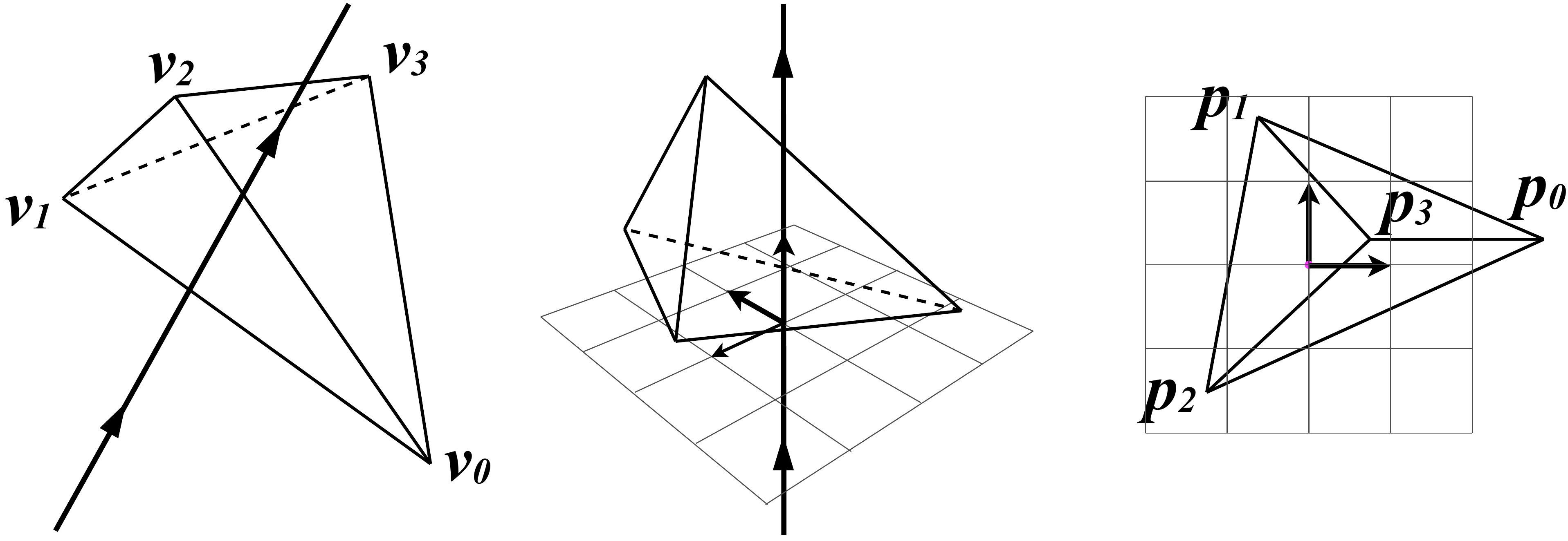}
\caption{Ray-tetrahedron intersection. Left: a ray and a tetrahedron. Middle: the tetrahedron transformed into the coordinate system defined by the ray. The ray coincides with the z-axis. Right: the tetrahedron projected onto 2-D. The ray is at the origin and points to the viewer.}
\label{fig:traversal_vis_annotated}
\end{figure}

Using the index $\textit{idx}_3$, we fetch the vertex from the points array, transform it to the new coordinate system, $C$, and use the resulting 2-D point $p_3$ to decide the exit face of the ray (cf.~Algorithm~\ref{alg:exit_face_selection}). Because the query point is at the origin after transformation, only four floating point multiplications and two floating point comparisons are sufficient. The exit face index is denoted as $\textit{exit}\_\textit{face}\_\textit{idx}$ and resides across the point $p_{\textit{exit}\_\textit{face}\_\textit{idx}}$ whose index is $\textit{idx}_{\textit{exit}\_\textit{face}\_\textit{idx}}$. To get the next tetrahedron, we use the $\textit{idx}_{\textit{exit}\_\textit{face}\_\textit{idx}}$ in the current tetrahedron data to fetch the corresponding neighbor tetrahedron index. Figure~\ref{fig:traversal_vis_annotated} illustrates the coordinate system transformation for a ray and a tetrahedron. 

\begin{algorithm}
  \caption{Exit face selection}
  \label{alg:exit_face_selection}
  \begin{algorithmic}
    \Procedure{GetExitFace}{$p_{0..3}$}
    \State $\textit{exit}\_\textit{face}\_\textit{idx} \gets 0$\;
    \If{$det(\vec{p_3}, \vec{p_0}) < 0$}
        \If{$det(\vec{p_3}, \vec{p_2}) \geq 0$}
        \State $\textit{exit}\_\textit{face}\_\textit{idx} \gets 1$\;
        \EndIf
    \ElsIf{$det(\vec{p_3}, \vec{p_1}) < 0$}
    \State $\textit{exit}\_\textit{face}\_\textit{idx} \gets 2$\;
    \EndIf
    \State \textbf{return} $\textit{exit}\_\textit{face}\_\textit{idx}$
    \EndProcedure    
  \end{algorithmic}
\end{algorithm}

In \textit{Tet32}, we simply search for $\textit{idx}_{\textit{exit}\_\textit{face}\_\textit{idx}}$ in the current tetrahedron. Since vertex and neighbor indices correspond to each other, location of the $\textit{idx}_{\textit{exit}\_\textit{face}\_\textit{idx}}$ (value from 0 to 3) also reveals the location of the neighbor to be traversed next. We describe this process in Algorithms~\ref{alg:traversal_tet32}~and~\ref{alg:get_next_tet32}.

\begin{algorithm}
  \caption{Tetrahedron traversal loop for \textit{Tet32}}
  \label{alg:traversal_tet32}
  \begin{algorithmic}
    \While{$\textit{tet}\_\textit{idx} \geq 0$}
    \State $\textit{idx}_{\textit{exit}\_\textit{face}\_\textit{id}} \gets \textit{idx}_3$\;
    \State $\textit{idx}_3 \gets \textit{idx}_0 \oplus \textit{idx}_1 \oplus \textit{idx}_2 \oplus \textit{VX}^{\textit{tet}\_\textit{idx}}$\;
    \State $v_\textit{new} \gets \textit{points}_{\textit{idx}_3} - {\textit{r}_o}$\;
    \State $p_{\textit{exit}\_\textit{face}\_\textit{idx}} \gets p_3$\;
    \State $p_3 \gets  (\vec{u} \cdot v_\textit{new}, \vec{v} \cdot v_\textit{new})$\;
    \State $\textit{exit}\_\textit{face}\_\textit{idx} = \Call{GetExitFace}{p_{0..3}}$
    \State $\textit{next}\_\textit{tet}\_\textit{idx} = \Call{GetNextTet32}{\textit{tet}\_\textit{idx}, \textit{idx}_i}$
    \EndWhile
  \end{algorithmic}
\end{algorithm}
\begin{algorithm}
  \caption{Next tetrahedron determination for \textit{Tet32}}
  \label{alg:get_next_tet32}
  \begin{algorithmic}
    \Procedure{GetNextTet32}{$\textit{tet}\_\textit{idx}, idx_{0..3}$}
    \State $\textit{next}\_\textit{tet}\_\textit{idx} \gets N_3^{\textit{tet}\_\textit{idx}}$
    \For{$i\gets 0, 2$}
      \If{$\textit{idx}_{\textit{exit}\_\textit{face}\_\textit{idx}} = V_i^{\textit{tet}\_\textit{idx}}$}
        \State $\textit{next}\_\textit{tet}\_\textit{idx} \gets N_i^{\textit{tet}\_\textit{idx}}$\;
      \EndIf
    \EndFor
    \State \textbf{return} $\textit{next}\_\textit{tet}\_idx$
    \EndProcedure    
  \end{algorithmic}
\end{algorithm}

In \textit{Tet20}, we use the property that neighbor indices are sorted using their counterpart vertex indices as keys. Thus, to find the next neighbor index, we find the order of $\textit{idx}_{\textit{exit}\_\textit{face}\_\textit{idx}}$ among $\textit{idx}_0$, $\textit{idx}_1$, $\textit{idx}_2$, $\textit{idx}_3$ (which are actually the vertex indices of the tetrahedron). Because the neighbor indices are sorted using vertex indices, order of the vertex index also happens to be the next neighbor index. We describe this process in Algorithms~\ref{alg:traversal_tet20}~and~\ref{alg:get_next_tet20}.

\begin{algorithm}
  \caption{Tetrahedron traversal loop for \textit{Tet20}}
  \label{alg:traversal_tet20}
  \begin{algorithmic}
    \While{$tet_{\textit{idx}} \geq 0$}
    \State $\textit{idx}_{\textit{exit}\_\textit{face}\_\textit{idx}} \gets \textit{idx}_3$
    \State $\textit{idx}_3 \gets \textit{idx}_0 \oplus \textit{idx}_1 \oplus \textit{idx}_2 \oplus \textit{VX}^{\textit{tet}\_\textit{idx}}$
    \State $v_\textit{new} \gets \textit{points}_{\textit{idx}_3} - {\textit{r}_o}$
    \State $p_{\textit{exit}\_\textit{face}\_\textit{idx}} \gets p_3$
    \State $p_3 \gets  (\vec{u} \cdot v_\textit{new}, \vec{v} \cdot v_\textit{new})$
    \State $\textit{exit}\_\textit{face}\_\textit{idx} = \Call{GetExitFace}{p_{0..3}}$
    \State $\textit{order}_a \gets$ sorted order of $id_{3}$ among $id_i$
    \State $\textit{next}\_\textit{tet}\_\textit{idx} = \Call{GetNextTet20}{\textit{tet}\_\textit{idx}, \textit{order}\_a}$
    \EndWhile
  \end{algorithmic}
\end{algorithm}
\begin{algorithm}
  \caption{Next tetrahedron determination for \textit{Tet20}}
  \label{alg:get_next_tet20}
  \begin{algorithmic}
    \Procedure{GetNextTet20}{$\textit{tet}\_\textit{idx}, \textit{order}\_a$}
    \State $\textit{next}\_\textit{tet}\_\textit{idx} \gets N_{\textit{order}\_a}$
    \State \textbf{return} $\textit{next}\_\textit{tet}\_idx$
    \EndProcedure    
  \end{algorithmic}
\end{algorithm}

\begin{algorithm}
  \caption{Tetrahedron traversal loop for \textit{Tet16}}
  \label{alg:traversal_tet16}
  \begin{algorithmic}
    \While{$\textit{tet}\_\textit{idx} \geq 0$}
    \State $\textit{idx}_3 \gets \textit{idx}_0 \oplus \textit{idx}_1 \oplus \textit{idx}_2 \oplus \textit{VX}^{\textit{tet}\_\textit{idx}}$\;
    \State $v_\textit{new} \gets \textit{points}_{\textit{idx}_3} - {\textit{r}_o}$
    \State $p_3 \gets  (\vec{u} \cdot v_\textit{new}, \vec{v} \cdot v_\textit{new})$\;
    \State $\textit{order}_a \gets$ sorted order of $id_{3}$ among $id_i$
    \State $\textit{exit}\_\textit{face}\_\textit{idx} = \Call{GetExitFace}{p_{0..3}}$
    \State $\textit{order}_b \gets$ sorted order of $id_{\textit{exit}\_\textit{face}\_\textit{idx}}$ among $id_i$
    \State $\textit{next}\_\textit{tet}\_\textit{idx} = $ 
    \State $\;\;\Call{GetNextTet16}{\textit{tet}\_\textit{idx}, \textit{prev}\_\textit{tet}\_\textit{idx}, \textit{order}\_a, \textit{order}\_b}$
    \State $\Call{Swap}{\textit{tet}\_\textit{idx}, \textit{prev}\_\textit{tet}\_\textit{idx}}$
    \EndWhile
  \end{algorithmic}
\end{algorithm}

\begin{algorithm}
  \caption{Next tetrahedron determination for \textit{Tet16}}
 \label{alg:get_next_tet16}
  \begin{algorithmic}
    \Procedure{GetNextTet16}{$\textit{tet}\_\textit{idx}, \textit{prev}\_\textit{tet}\_\textit{idx}, \newline 
                   \hspace*{2cm} \textit{order}\_a, \textit{order}\_b$}
    \If{$\textit{order}_a \neq 3$}
      \State $\textit{next}\_\textit{tet}\_\textit{idx} = \textit{prev}\_\textit{tet}\_\textit{idx} \oplus \textit{NX}_{\textit{order}_a}^{\textit{tet}\_\textit{idx}}$
    \EndIf
    \If{$\textit{order}_b \neq 3$}
      \State $\textit{next}\_\textit{tet}\_\textit{idx} = \textit{next}\_\textit{tet}\_\textit{idx} \oplus \textit{NX}_{\textit{order}_b}^{\textit{tet}\_\textit{idx}}$
    \EndIf
    \State \textbf{return} $\textit{next}\_\textit{tet}\_\textit{idx}$
    \EndProcedure    
  \end{algorithmic}
\end{algorithm}

In \textit{Tet16}, we use the previous tetrahedron index to reconstruct next tetrahedron index using the values $\textit{NX}^i_j$. As in $Tet20$, we need to construct the value $\textit{NX}^i_j$ using sorted vertex indices. To reconstruct the next tetrahedron, sorted order of values are computed for $\textit{idx}_3$ which corresponds to a previous tetrahedron and $\textit{idx}_{\textit{exit}\_\textit{face}\_\textit{idx}}$, which corresponds to an exit face, must be computed. We describe this process in Algorithms~\ref{alg:traversal_tet16}~and~\ref{alg:get_next_tet16}. 

If the neighbor index points to a constrained face or tetrahedral mesh boundaries, we terminate the traversal. Otherwise, knowing the next tetrahedron, we discard $p_i$ and $\textit{idx}_i$ by replacing its contents with the newly fetched point data $p_3$ and $\textit{idx}_3$. We repeat this process until a geometry is intersected or the tetrahedral mesh boundaries are reached. In this method, no further modifications are necessary to ensure clockwise ordering because the counterclockwise ordering is always preserved for points on the exit face.

Fetching a new vertex id requires three bitwise exclusive-or operations. The coordinate system transformation of the newly fetched point is six floating-point multiplications and four floating-point additions. We decide whether a face is an exit face by using four floating-point multiplications and two floating-point comparisons. Finally, we determine the next tetrahedron index using the appropriate method for the preferred structure.

\subsection{Point projection using specialized basis}
\label{subsection:PointProj}
We project newly fetched points to the 2-D coordinate system using two dot product operations, which require six floating-point multiplications and four floating-point additions. We can optimize this step by scaling the basis vectors to make some of the components zero or one. Since the basis vectors are only scaled, the exit face determination still works correctly. To avoid numerical issues, we scale vectors in such a way that only the absolute largest components become one (or minus one). Equation~(\ref{eq:basis1}) describes the construction of the first basis vector $\vec{u}$, which is orthogonal to $\vec{n}$ (and not necessarily of unit length). 

\begin{equation}
\begin{aligned}
\vec{u}_{\textit{min}} ={} & 0, \\
\vec{u}_{(\textit{min}+1)\ \textit{mod}\ 3} ={} & \dfrac{\vec{n}_{(\textit{min}-1)\ \textit{mod}\ 3}}{\vec{n}_{\textit{max}}},   \\
\vec{u}_{(\textit{min}-1)\ \textit{mod}\ 3} ={} & -\dfrac{\vec{n}_{(\textit{min}+1)\ \textit{mod}\ 3}}{\vec{n}_{\textit{max}}},
\end{aligned}
\label{eq:basis1}
\end{equation}

\noindent where $\vec{v}_0$, $\vec{v}_1$, and $\vec{v}_2$ correspond to $\vec{v}_x$, $\vec{v}_y$, and $\vec{v}_z$, respectively, \textit{min} and \textit{max} are the indices of the absolute smallest and largest components of the vector $\vec{n}$.

We construct the second basis vector $\vec{v}$, which is orthogonal to $\vec{n}$ and $\vec{u}$ (and not necessarily of unit length), as in Equation~(\ref{eq:basis2}).

\begin{equation}
\begin{aligned}
\vec{t} ={} & \vec{n} \times \vec{u}, \\
\vec{v} ={} & \dfrac{\vec{t}}{\vec{t}_{(3-\textit{max}-\textit{min})}}.
\end{aligned}
\label{eq:basis2}
\end{equation}

Now, we can transform 3-D point $v$ to the 2-D coordinate system using the basis $b = (\vec{u}, \vec{v})$, as shown in Equation~\ref{eq:basis_usage}. It should be noted that the sign $s$ of the last parameter $\vec{v}_{\textit{min}}$ can be either positive or negative depending on the sign of $\vec{v}_{\textit{min}}$. 

\begin{equation}
\begin{aligned}
\textit{other} ={} & 3 - \textit{max} - \textit{min}, \\
\vec{p}_x ={} & \vec{u}_{\textit{max}} \vec{v}_{\textit{max}} + \vec{v}_{\textit{other}}, \\
\vec{p}_y ={} & \vec{v}_{\textit{max}} \vec{v}_{\textit{max}} + \vec{v}_{\textit{other}} \vec{v}_{\textit{other}} \pm \vec{v}_{\textit{min}}.
\end{aligned}
\label{eq:basis_usage}
\end{equation}

To perform the above computation, three floating-point multiplications and three floating-point addition/subtractions are sufficient. We implement this fast projection method using a templated function over the variables \textit{min}, \textit{max}, and \textit{sign}$(\vec{v}_{\textit{min}})$ and call the corresponding function by inspecting the components of the new basis to avoid run-time overhead of keeping additional function arguments.

\subsection{Locating points in a tetrahedral mesh}
\label{sec:locating_points}
To initiate the ray traversal, the tetrahedron that contains the common ray origin has to be located. Similarly, to cast shadow rays, we need to identify the tetrahedra that contain light sources. For this purpose, we use the 3-D variant of the proposed tetrahedral mesh traversal methods described earlier using a predetermined source tetrahedron as the first tetrahedron for traversal. 

\section{Reordering tetrahedral mesh data}

\begin{figure}[h!]
  \includegraphics[width=\columnwidth]{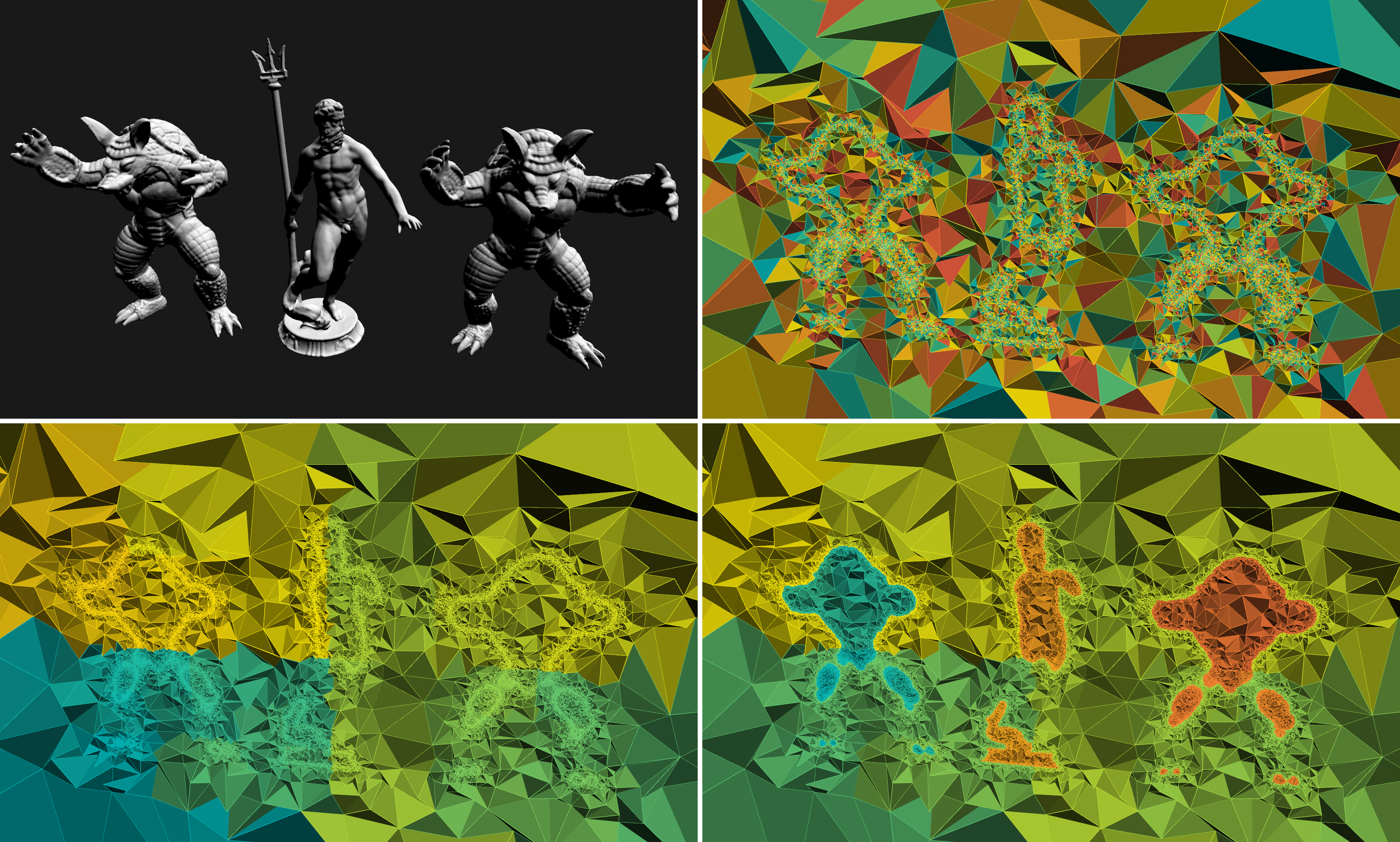}
  \caption{Sorting tetrahedron data. Top left: The three-dimensional scene. Top right: Unsorted tetrahedron data. Bottom left: Tetrahedron data sorted using an Hilbert curve. Bottom right: Tetrahedron data sorted using an Hilbert curve and mesh regions. Memory positions are coded with different colors.}
  \label{sorting_tetrahedra}
\end{figure}

We reorder points and tetrahedra in memory to improve cache locality during ray-traversal. For this purpose, we use a two-step method. In the first step, we detect if there are distinct regions in the tetrahedralization. These regions occur when a set of tetrahedra is completely enclosed by a set of constrained faces. Because the rays are traced until a constrained face is encountered, the tetrahedra from different regions are not visited in a single ray traversal, which is not the case for multi-hit traversal methods. Thus, we store the tetrahedra that belong to the same region close together in memory. Furthermore, we reorder points based on their positions and tetrahedra based on their center points. We map points to memory using a Hilbert curve (see Figure~\ref{sorting_tetrahedra}, bottom left). Hilbert curve is a space-filling curve that can be used to map spatial data from 3-D to 1-D by preserving the locality. This means that primitives that are close to each other in 3-D space are also close to each other in 1-D. 

\section{Handling Common Ray-tracing Operations}

\begin{figure}[h!]
\includegraphics[width=\columnwidth]{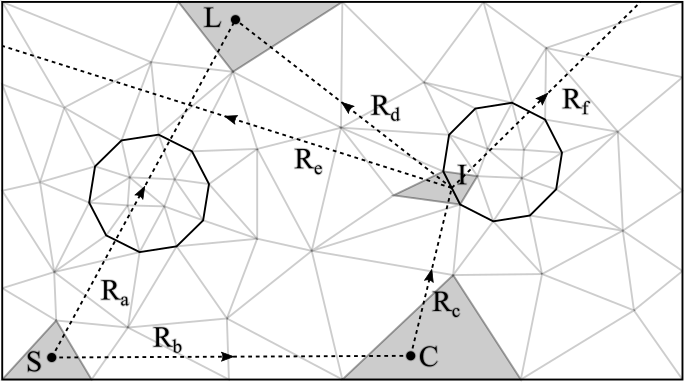}
\caption{The types of rays in tetrahedral mesh-based ray tracing. $S$ is the center of source tetrahedron. $C$ is the camera position. $I$ is the intersection point. $L$ is the light source position. Rays $R_a$ and $R_b$ are used to locate the light source and camera position, respectively. $R_c$ is the camera ray. $R_d$ is the shadow ray. $R_e$ and $R_f$ are the reflection and refractions rays, which are cast from two different neighboring tetrahedra.}
\label{fig:different_rays}
\end{figure}

We handle common ray-tracing operations using tetrahedral meshes as follows. Handling mesh lights is straightforward by using the proposed traversal methods. For point lights, we locate the tetrahedron that contains the point light at the start of each frame. Then, we use a slightly modified traversal algorithm where the traversal terminates if the tetrahedron that contains the light source is reached. We cast reflection and refraction rays using the neighboring tetrahedron on the shared face of the tetrahedron in which the traversal is terminated. In this way, we avoid an intersection with the same face. To handle shadow, reflection, and refraction rays together, we report the two tetrahedra that share the common intersected face in the intersection routine. Figure~\ref{fig:different_rays} illustrates different types of rays used in a tetrahedral mesh-based ray tracing. At the start of each frame, we locate the camera and the point light sources and store their tetrahedron indices. For this purpose, we start from a source tetrahedron $S$ that can be arbitrarily chosen and locate the tetrahedra that contain the camera and the point light sources.  

\section{GPU Implementation}
For the GPU implementation, we use the CUDA platform. Once we build the tetrahedral mesh-based acceleration structure, the tetrahedra and points data are copied to the GPU. We store the constrained face data on the host computer because it is not a part of the \textit{hot} data, which is frequently accessed during traversal. Once initialization is complete, the steps to render a single frame are as follows.
\begin{enumerate}
    \item We identify the \textit{source tetrahedron} on the CPU, as described in Section~\ref{sec:locating_points}
    \item We pass the batch of rays and the source tetrahedron to the global memory of the GPU.
    \item CUDA kernels run for each ray, traversing the scene, and terminate when they hit the scene geometry.
    \item We store the results of the intersection calculations in the global memory of the GPU and then passed them to the main memory. We then use these results to perform shading and to generate additional rays.
\end{enumerate}

Our method can be trivially implemented for the CUDA platform. However, this trivial implementation does not provide the best performance on the GPU in terms of computation speed. Thus, we perform the following optimizations to make our method run faster on the GPU.
\begin{enumerate}
    \item We project ray origin to the 2-D coordinate system beforehand. When projecting the newly fetched point, translation is performed on the 2-D coordinate system instead of a 3-D one. Thus, instead of using the origin in 3-D, we use projected origin in 2-D. This potentially results in fewer occupied registers on the GPU, resulting in better performance. We compute the projected origin, $\textit{po}$, as follows: 
    \begin{equation}
    \textit{po} = (\vec{u} \cdot r_o, \vec{v} \cdot r_o),
    \label{eq:projected_origin_1}
    \end{equation}
    \noindent where $(\vec{u}, \vec{v})$ is the 2-D basis constructed from the ray.
    During traversal, we can project the new point to the 2-D plane as follows: 
    \begin{equation}
    p_3 = (\vec{u} \cdot v_{new} - po_x, \vec{v} \cdot v_{new} - po_y),
    \label{eq:projected_origin_2}
    \end{equation}
    \noindent where $p_3$ is the projected point and $v_{\textit{new}}$ is the newly fetched point from the next tetrahedron.
    \item We make use of CUDA textures when accessing tetrahedral mesh data. To optimize traversal in \textit{Tet20} and \textit{Tet16} structures, we use a single channel integer (32 bytes) texture. To use it, the required elements are the xor field and one neighbor field for \textit{Tet20}, the xor field and one or two neighbor fields for \textit{Tet16}). We fetch and store these in the local stack; potentially reducing the maximum number of registers used.
\end{enumerate}

\section{Experimental Results}
We compare our approach to \textit{k}-d trees, BVHs, and the state-of-the-art tetrahedral mesh-based methods, namely the ScTP-based traversal~\cite{Lagae:2008:Constrained} and the Pl\"{u}cker coordinate-based traversal~\cite{Maria:2017:Traversal}. We use the \textit{k}-d tree and SAH-based BVH implementations, as described in~\cite{Wald:2007:FastSAH} and~\cite{Gunther:2007:Packet}. We use the original implementation provided by Maria et al.~\cite{Maria:2017:Traversal} for the Pl\"{u}cker coordinate-based traversal. 

We use TetGen~\cite{Si:2015:TetGen} to generate the tetrahedral mesh of the 3D scene. We perform experiments on a computer with six cores @3.2 GHz (Intel), 16 GB of main memory, and NVIDIA GTX 1060 with 6 GB of memory. On the CPU, we render the scenes using multi-threading by subdividing the image into 16$\times$16 tiles and assigning them to available threads. We render the images at 1920$\times$1440 resolution. To make a fair comparison between our method and the other state-of-the-art approaches, we render the same scene many times and pick the best result for each method to avoid noisy measurements due to background processes.

Tables~\ref{tab:build_and_render_times}~and~\ref{tab:build_and_render_times_remesh} show the computational costs of  the construction of acceleration structures and rendering times of different traversal methods for test scenes. The test scenes in Table ~\ref{tab:build_and_render_times_remesh} cannot be tetrahedralized using TetGen. Therefore, we tetrahedralized them using TetWild~\cite{Hu:2018:TetWild} and used the remeshed geometry produced by TetWild as an input geometry. To test the adaptiveness of the structures in a challenging scene geometry, we include the versions of the scenes with bounding boxes composed of large triangles. Experiments show that our method performs better than the ScTP-~\cite{Lagae:2008:Constrained} and Pl\"{u}cker coordinate-based traversal methods~\cite{Maria:2017:Traversal} in all test scenes. It performs better than the BVH-based traversal in seven of the fifteen scenes and better than the \textit{k}-d tree-based traversal in six of the fifteen scenes. In the other test scenes that BVH- and \textit{k}-d tree-based traversal methods perform superior to our tetrahedral mesh-based traversal, the rendering times are mostly close to each other. While testing the state-of-the-art tetrahedral mesh-based traversal methods of~\cite{Lagae:2008:Constrained} and~\cite{Maria:2017:Traversal}, we sorted the tetrahedral meshes using space-filling curves for a fair comparison. Although the construction times of BVHs and \textit{k}-d trees are lower than that of the tetrahedral meshes, the tetrahedral mesh is constructed during preprocessing and it does not affect the raytracing performance for the scenes that do not require the update of acceleration structures. The tetrahedral mesh does not need to be updated for dynamic scenes where the topology does not change. If the topological changes to a tetrahedralization are local, the tetrahedral mesh can be updated with efficient insertion and removal operations~\cite{Lagae:2008:Constrained}. 

Table~\ref{tab:render_times_gpu} shows rendering times of different tetrahedral mesh-based methods for test scenes on the GPU. \textit{Tet20} representation gives the best performance. It is around 15\% faster than Maria's method while occupying much less memory (half of the memory required by Maria's method in the largest test scene). \textit{Tet16} representation requires even less memory but it is not as fast as \textit{Tet20} (roughly the same performance as Maria's method) due to more memory and arithmetic operations needed to decode compressed neighbor data.

Table~\ref{tab:memory_usage} shows the memory costs for different acceleration structures on different scenes. Our most compact structure, TetMesh16, can be stored using significantly less memory than the other alternatives, which provides two benefits. First, accelerators for much larger scenes can be fitted to the main memory or GPU global memory. Second, this small footprint provides much better performance by facilitating cache locality. It should be noted that our smallest accelerator data is memory aligned (16 bytes per each tetrahedron).

Figure~\ref{fig:tet_mesh_sorting} demonstrates the effect of the tetrahedral mesh sorting on rendering performance. Even though sorting is not vital for performance in small scenes, it significantly improves the rendering performance in large scenes. 
\input{tet_mesh_sorting/figure.tex}

Table~\ref{tab:camera_distance} demonstrates the efficiency of a tetrahedral mesh-based traversal approach when the camera gets closer to a surface. In this experiment, we render the images at varying distances to the 3-D model of the Armadillo and compare the rendering times for different acceleration structures. Both BVH and \textit{k}-d tree performs much better than the tetrahedral mesh structure when the camera views the object from a fair distance. However, as the camera gets closer to a surface, the traversal cost decreases because the tetrahedral mesh structure is not hierarchical, unlike the BVH and \textit{k}-d tree. In the extreme case, when the camera is about to touch the surface, only one tetrahedron is traversed. This is not the case for hierarchical structures because many tree nodes may need to be traversed to find the closest ray-surface intersection.

Although representations proposed by Lagae and Dutre~\cite{Lagae:2008:Constrained} and Maria et al.~\cite{Maria:2017:Traversal} and our method use similar mesh representations, the performance difference between them is because of the following reasons:
\begin{itemize}
    \item \textit{Memory operations per tetrahedron:} We only fetch one point per tetrahedron, thanks to the xor-based storage scheme. In Tet-mesh-ScTP and Pl\"{u}cker based method, all four points are fetched from the memory. Although three of them will be in the cache because three points are shared between tetrahedra, it still costs more than fetching only one point.
    \item \textit{Compact storage:} Our method requires less memory than the approaches we compare. This speeds up the computations because the cache utilization is high. This also allows us to render larger scenes since more geometry can be fitted to the memory.
    \item \textit{Arithmetic operations per tetrahedron:} Tet-mesh-ScTP relies on a scalar triple product, which accounts for 40 floating points on average for the computation of the terms. Similarly, the method proposed by Maria et al.~\cite{Maria:2017:Traversal} also works in 3-D, thus resulting in more expensive computations. On the other hand, our transformed 2-D coordinate system results in very few arithmetic operations (13 floating-point operations). Because we project points as soon as they are fetched from the memory, they occupy few registers. In Tet-mesh-ScTP, there may be a possible performance loss due to more register usage.
    \item \textit{Determining the next tetrahedron:} In our method, we never take the previous tetrahedron into account as the next tetrahedron to visit (similar to Maria~et~al.~\cite{Maria:2017:Traversal}). However, Tet-mesh-ScTP takes all four neighbors into account by computing 3-6 scalar triple products to determine the next tetrahedron, which makes the computations more costly. Besides, this may reduce the effectiveness of the branch prediction as well because there are more candidate neighbors.
\end{itemize}

\section{Conclusions and Future Research Directions}

We propose methods for fast tetrahedral mesh traversal for ray tracing. Specifically, we propose a compact and memory-aligned tetrahedral mesh data structure. We use a space-filling curve to improve cache locality. We propose an efficient traversal method to improve ray-tracing performance and provide its GPU implementation. Experiments show that our approach can reduce rendering times substantially and perform better than other alternatives in different scenarios. There are two main limitations of using tetrahedral meshes as acceleration structures in ray tracing complex three-dimensional scenes.

\begin{itemize}
    \item Tetrahedral mesh generation process is computationally costly and requires a significant amount of memory than the alternative methods.
    \item Our current implementation is not able to construct a tetrahedral mesh acceleration structure for scenes with intersecting geometry. We can overcome this limitation by a pre-processing step where mesh intersections are resolved so that the resulting geometry is a Piecewise Linear Complex (PLC)~\cite{Miller:1998:VolumeMeshes}, which is proposed in~\cite{Lagae:2008:Constrained}.
\end{itemize}

Other areas for further research of our tetrahedral mesh-based acceleration structure regarding contemporary ray-tracing concepts are as follows.
\begin{itemize}
    \item \textit{Instancing:} In its current form, our method cannot easily handle instances. However, the strengths of tetrahedral-mesh based accelerators can still be utilized if tetrahedral-meshes are built per model and shared among the instances given that these accelerators are put into the nodes of another acceleration structure like a BVH.
\end{itemize}    

\begin{itemize}
    \item \textit{Non-triangular models:} The proposed acceleration structure does not support non-triangular models. Recent research by Hu et al.~\cite{Hu:2019:TriWild} provide a way to build triangulations with curve constraints. The extension of this method to 3-D with surface constraints can act as an accelerator, which could be a potentially interesting and challenging research direction.
    \item \textit{Real-time rebuilds:} Although our approach allows real-time manipulation of the geometry by certain deformers (smooth, C1 continuous) naturally, it is not very easy to have real-time rebuilds on changing geometry, which is well supported by the state of the art BVHs. 
\end{itemize}

We plan to experiment with the triangulations with curve constraints~\cite{Hu:2019:TriWild}. The extension of this method to 3-D would allow us to render parametric 3-D surfaces directly using tetrahedralizations.

Another potential use case is volume visualization. Even though the adaptation of our approach to direct volume rendering would result in a slower traversal (and possibly overlap with the approach employed by Marmitt et al.~\cite{Marmitt:2006:Traversal}), there are still two potential improvements it can provide:
\begin{itemize}
    \item[{\textit{i)}}] First, our compact structure would result in better cache utilization, thereby reducing the computation time.
    \item[{\textit{ii)}}] Second, this compact structure would need less memory and enable visualization of larger models that can fit into the memory. This is even more critical in GPU, where the memory is relatively limited.
\end{itemize}

\section{Acknowledgments}
This research is supported by The Scientific and Technological Research Council of Turkey (T\"{U}B\.{I}TAK) under Grant No. 117E881. We are grateful to Dr. Maxime Maria and his colleagues for providing us their implementation of the tetrahedral mesh traversal method.

\input{build_and_render_times/table_black}
\input{render_times_gpu/table_black.tex}
\input{memory_usage/table_black.tex}
\input{camera_distance/table.tex}

\clearpage



\end{document}

%% file: tet_mesh_sorting/figure.tex
\begin{figure}
{\small
\begin{tikzpicture}
\begin{axis}[
ybar = 0.6,
ymin = 0,
yticklabel style={/pgf/number format/fixed, /pgf/number format/precision=5},
scaled y ticks=false,
enlarge x limits=0.2,
enlarge y limits={upper, value=0.2},
ylabel = Rendering time (sec.),
legend style = { at = { (0.5,0.95) },
anchor = north,legend columns = -1 },
width = \columnwidth,
xticklabel style={rotate=20},
symbolic x coords={Torus Knots, Armadillo, Neptune, Mix, Mix Close, },
xtick = data,
]
\addplot
coordinates{
(Torus Knots,0.248)
(Armadillo,0.242)
(Neptune,0.259)
(Mix,0.468)
(Mix Close,0.474)
};
\addplot
coordinates{
(Torus Knots,0.221)
(Armadillo,0.204)
(Neptune,0.22)
(Mix,0.371)
(Mix Close,0.395)
};
\addplot
coordinates{
(Torus Knots,0.216)
(Armadillo,0.207)
(Neptune,0.222)
(Mix,0.372)
(Mix Close,0.39)
};
\legend{
None, Hilbert, Hilbert-regions, }
\end{axis}
\end{tikzpicture}
}
\caption{The rendering times for unsorted and sorted tetrahedral mesh data.}
\label{fig:tet_mesh_sorting}
\end{figure}
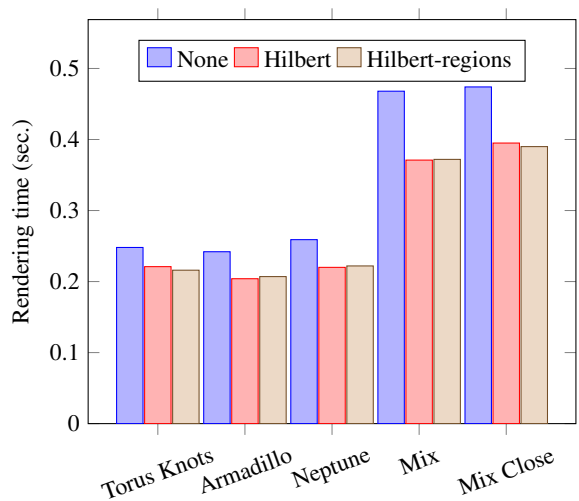

%% file: build_and_render_times/table_black.tex
\begin{table*}[htbp]
\centering
\small
\caption{The computational costs of acceleration structures and rendering times for traversal methods}
\begin{tabular}{|lrrrrr|}
\hline
\multicolumn{6}{|c|}{Scenes}\\
\hline
& \multicolumn{1}{c}{Torus Knots}
& \multicolumn{1}{c}{Torus Knots in a Box}
& \multicolumn{1}{c}{Armadillo}
& \multicolumn{1}{c}{Armadillo in a Box}
& \multicolumn{1}{c|}{Neptune}
\\
& \includegraphics[width = 0.14\textwidth]{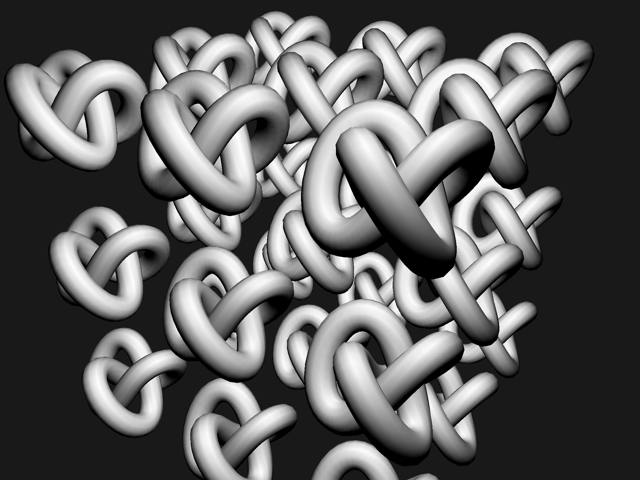} & \includegraphics[width = 0.14\textwidth]{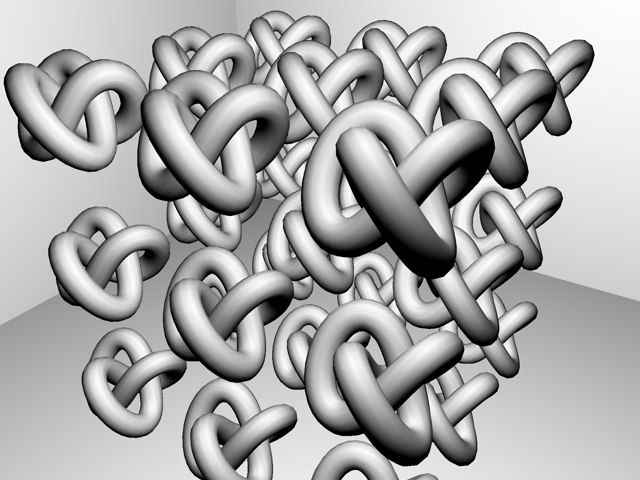} & \includegraphics[width = 0.14\textwidth]{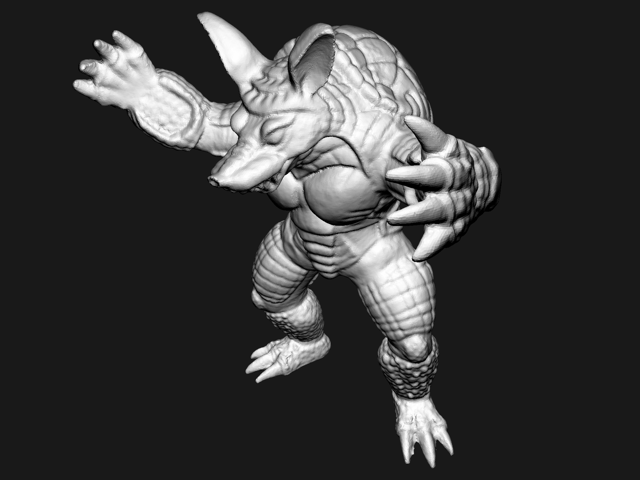} & \includegraphics[width = 0.14\textwidth]{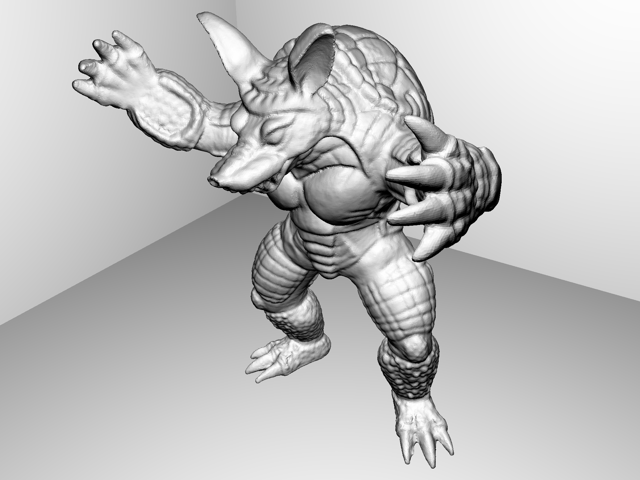} & \includegraphics[width = 0.14\textwidth]{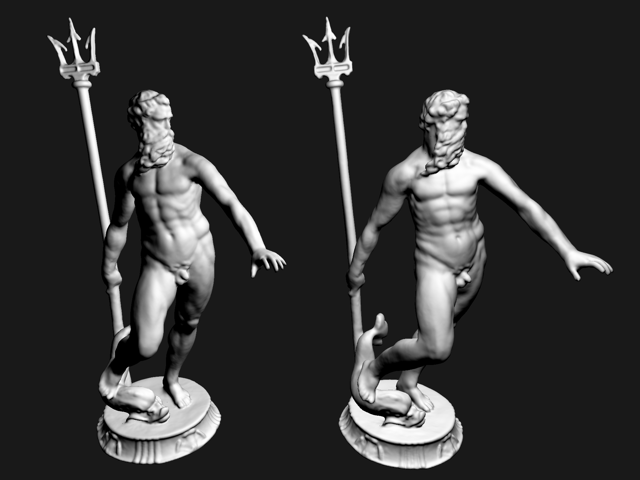} \\
\hline
\hline
\multicolumn{6}{|c|}{Scene statistics} \\
\hline
\# of triangles & 77,760 & 77,772 & 345,938 & 345,950 & 448,896 \\
\hline
\hline
\multicolumn{6}{|c|}{Construction times (in seconds)} \\
\hline
Tet-mesh-ScTP~\cite{Lagae:2008:Constrained}& 3.596 & 3.638 & 16.733 & 20.252 & 79.822 \\
Tet-mesh-80~\cite{Maria:2017:Traversal}& 3.656 & 3.663 & 16.595 & 20.143 & 79.657 \\
Tet-mesh-32& 3.753 & 3.643 & 16.659 & 20.262 & 79.536 \\
Tet-mesh-20& 3.778 & 3.658 & 16.704 & 20.444 & 79.573 \\
Tet-mesh-16& 3.640 & 3.524 & 16.546 & 19.919 & 79.846 \\
BVH~\cite{Matt:2016:Pbr}& \textbf{0.078} & \textbf{0.079} & \textbf{0.391} & \textbf{0.396} & \textbf{0.474} \\
\textit{k}-d tree~\cite{Matt:2016:Pbr}& 0.739 & 0.590 & 1.454 & 1.651 & 2.265 \\
\hline
\hline
\multicolumn{6}{|c|}{Rendering times (in milliseconds)} \\
\hline
Tet-mesh-ScTP~\cite{Lagae:2008:Constrained}& 261.5 & 293.4 & 232.5 & 306.7 & 268.9 \\
Tet-mesh-80~\cite{Maria:2017:Traversal}& 244.1 & 278.9 & 218.1 & 262.4 & 236.5 \\
Tet-mesh-32& 150.7 & 181.7 & 148.5 & 182.5 & 158.7 \\
Tet-mesh-20& \textbf{125.8} & \textbf{142.3} & 117.1 & 145.3 & 127.1 \\
Tet-mesh-16& 136.3 & 152.4 & 124.6 & 153.2 & 135.9 \\
BVH~\cite{Matt:2016:Pbr}& 152.7 & 192.2 & \textbf{78.1} & \textbf{126.1} & \textbf{78.7} \\
\textit{k}-d tree~\cite{Matt:2016:Pbr}& 139.9 & 214.4 & 85.7 & 182.3 & 81.7 \\
\hline
\multicolumn{6}{c}{}\\
\hline
\multicolumn{6}{|c|}{Scenes}\\
\hline
& \multicolumn{1}{c}{Neptune in a Box}
& \multicolumn{1}{c}{Mix}
& \multicolumn{1}{c}{Mix in a Box}
& \multicolumn{1}{c}{Mix close}
& \multicolumn{1}{c|}{Mix in a Box close}
\\
& \includegraphics[width = 0.14\textwidth]{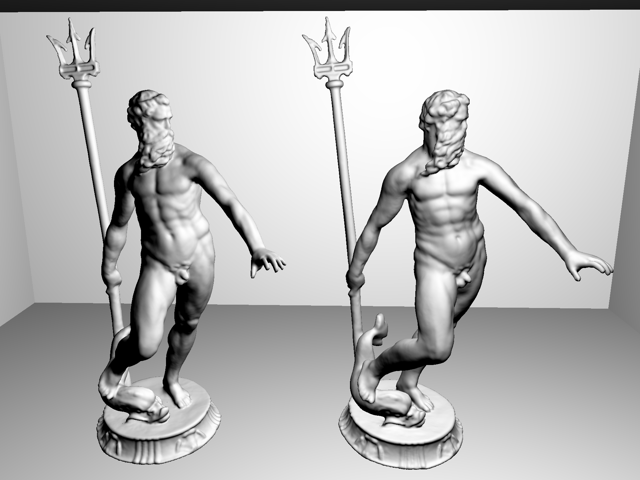} & \includegraphics[width = 0.14\textwidth]{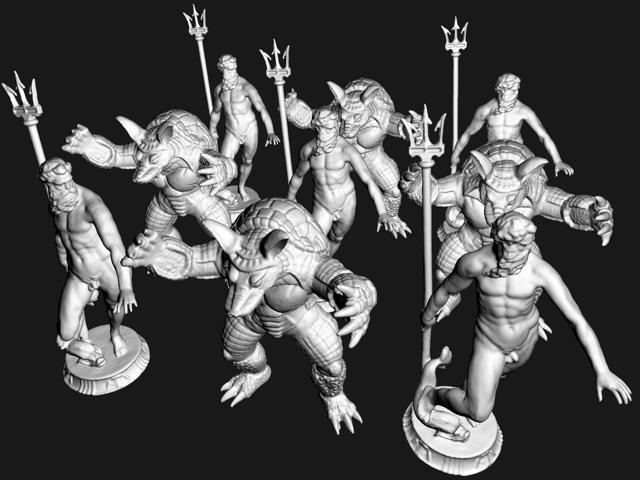} & \includegraphics[width = 0.14\textwidth]{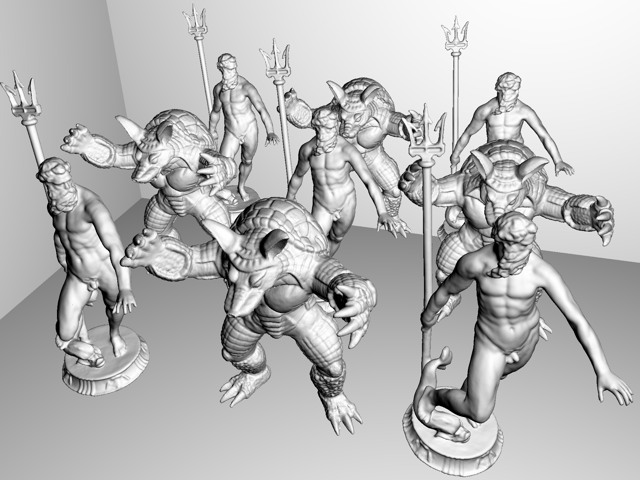} & \includegraphics[width = 0.14\textwidth]{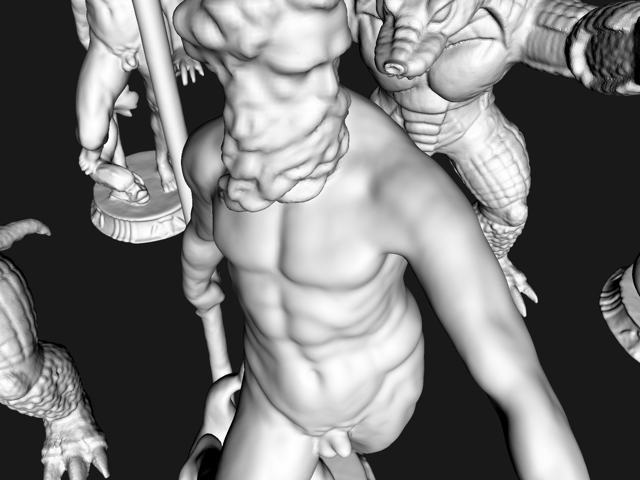} & \includegraphics[width = 0.14\textwidth]{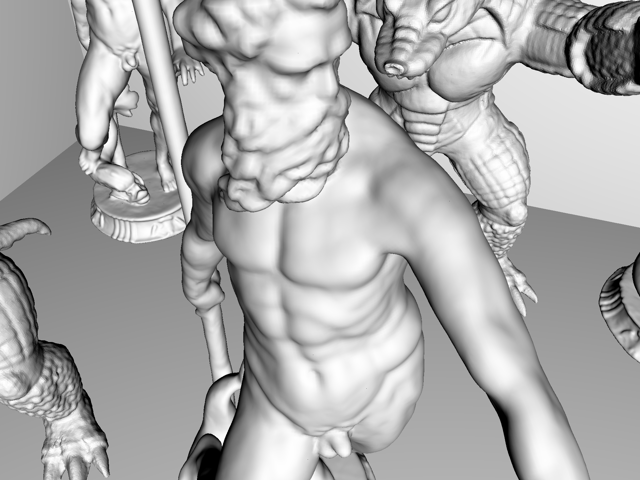} \\
\hline
\multicolumn{6}{c}{}\\
\hline
\multicolumn{6}{|c|}{Scene statistics} \\
\hline
\# of triangles & 448,908 & 2,505,992 & 2,506,004 & 2,505,992 & 2,506,004 \\
\hline
\multicolumn{6}{c}{}\\
\hline
\multicolumn{6}{|c|}{Construction times (in seconds)} \\
\hline
Tet-mesh-ScTP~\cite{Lagae:2008:Constrained}& 155.478 & 124.208 & 170.216 & 124.840 & 485.950 \\
Tet-mesh-80~\cite{Maria:2017:Traversal}& 156.381 & 125.081 & 169.116 & 125.068 & 482.908 \\
Tet-mesh-32& 153.788 & 124.000 & 169.502 & 123.183 & 487.291 \\
Tet-mesh-20& 155.580 & 124.087 & 169.890 & 124.015 & 483.827 \\
Tet-mesh-16& 154.644 & 124.493 & 170.175 & 123.401 & 484.516 \\
BVH~\cite{Matt:2016:Pbr}& \textbf{0.487} & \textbf{2.968} & \textbf{3.017} & \textbf{2.966} & \textbf{2.997} \\
\textit{k}-d tree~\cite{Matt:2016:Pbr}& 2.471 & 13.889 & 14.668 & 13.846 & 16.624 \\
\hline
\multicolumn{6}{c}{}\\
\hline
\multicolumn{6}{|c|}{Rendering times (in milliseconds)} \\
\hline
Tet-mesh-ScTP~\cite{Lagae:2008:Constrained}& 279.6 & 402.6 & 430.0 & 449.5 & 455.0 \\
Tet-mesh-80~\cite{Maria:2017:Traversal}& 261.0 & 355.7 & 384.7 & 411.2 & 419.6 \\
Tet-mesh-32& 176.1 & 247.2 & 268.5 & 265.5 & 269.9 \\
Tet-mesh-20& 137.0 & 196.3 & 211.1 & 205.6 & \textbf{210.2} \\
Tet-mesh-16& 152.0 & 223.9 & 241.4 & 237.4 & 240.3 \\
BVH~\cite{Matt:2016:Pbr}& \textbf{120.4} & 144.6 & \textbf{187.9} & 224.9 & 253.8 \\
\textit{k}-d tree~\cite{Matt:2016:Pbr}& 162.6 & \textbf{143.5} & 214.8 & \textbf{193.1} & 213.6 \\
\hline
\end{tabular}
\label{tab:build_and_render_times}
\end{table*}

\begin{table*}[htbp]
\centering
\small
\caption{The computational costs of  acceleration structures and rendering times for  traversal methods (remeshed scenes).}
\begin{tabular}{|lrrrrr|}
\hline
\multicolumn{6}{|c|}{Scenes}\\
\hline
& \multicolumn{1}{c}{Rungholt Far}
& \multicolumn{1}{c}{Rungholt Default}
& \multicolumn{1}{c}{Rungholt Close}
& \multicolumn{1}{c}{Exhaust Pipe Left}
& \multicolumn{1}{c|}{Exhaust Pipe Right}
\\
& \includegraphics[width = 0.14\textwidth]{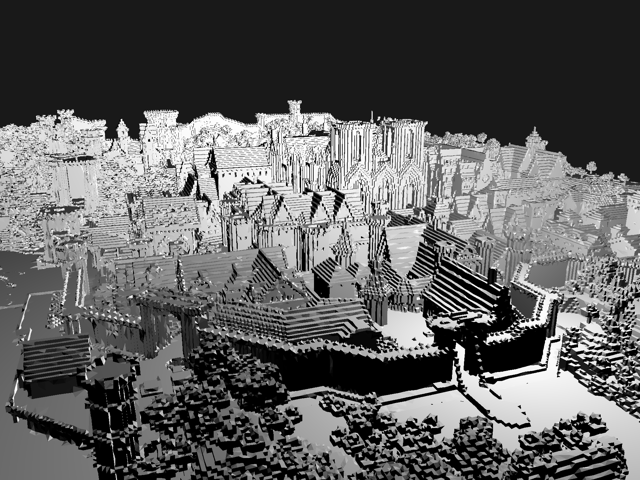} & \includegraphics[width = 0.14\textwidth]{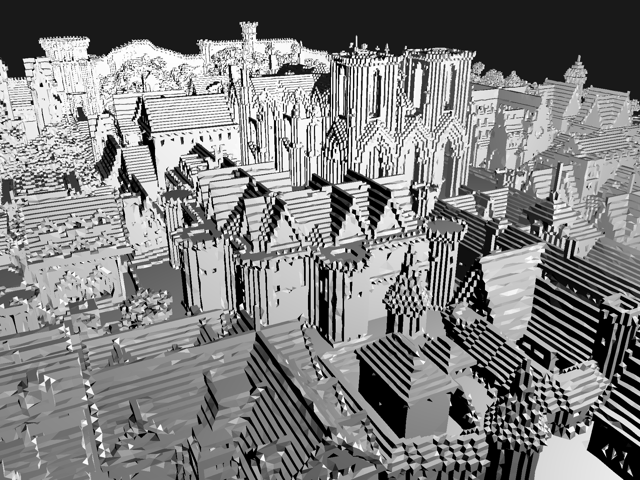} & \includegraphics[width = 0.14\textwidth]{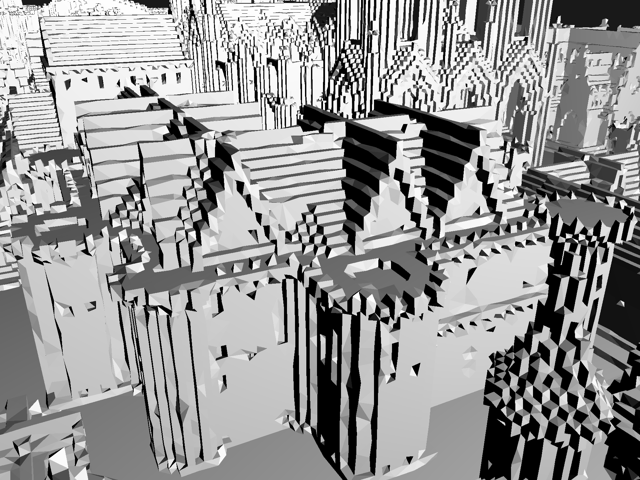} & \includegraphics[width = 0.14\textwidth]{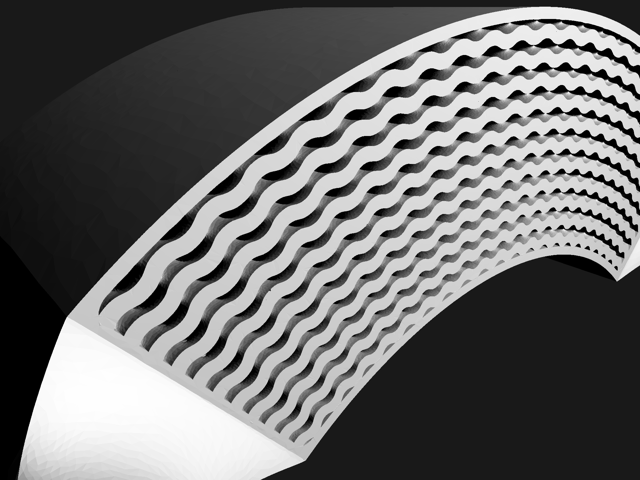} & \includegraphics[width = 0.14\textwidth]{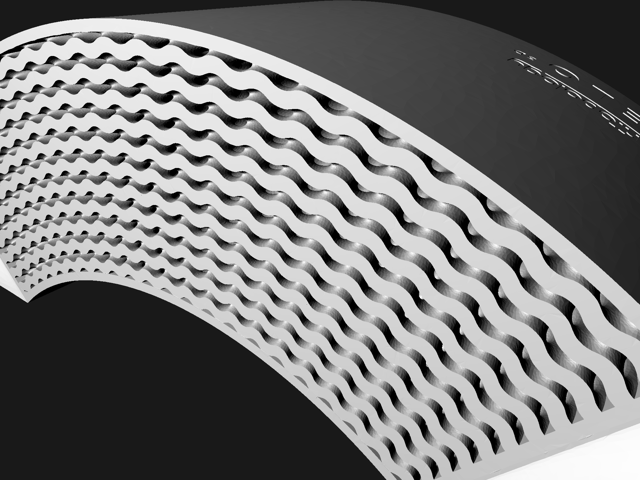} \\
\hline
\multicolumn{6}{c}{}\\
\hline
\multicolumn{6}{|c|}{Scene statistics} \\
\hline
\# of triangles & 3,580,928 & 3,580,928 & 3,580,928 & 6,244,678 & 6,244,678 \\
\hline
\multicolumn{6}{c}{}\\
\hline
\multicolumn{6}{|c|}{Construction times (in seconds)} \\
\hline
BVH~\cite{Matt:2016:Pbr}& 4.230 & 4.247 & 4.212 & 7.771 & 7.767 \\
\textit{k}-d tree~\cite{Matt:2016:Pbr}& 37.140 & 37.078 & 37.039 & 66.462 & 66.403 \\
\hline
\multicolumn{6}{c}{}\\
\hline
\multicolumn{6}{|c|}{Rendering times (in milliseconds)} \\
\hline
Tet-mesh-ScTP~\cite{Lagae:2008:Constrained}& 554.035 & 525.523 & 436.716 & 333.291 & 344.483 \\
Tet-mesh-80~\cite{Maria:2017:Traversal}& 488.299 & 466.933 & 400.589 & 312.152 & 320.361 \\
Tet-mesh-32& 353.444 & 333.274 & 265.337 & 202.472 & 207.239 \\
Tet-mesh-20& 282.211 & 265.337 & 215.118 & 163.814 & 166.619 \\
Tet-mesh-16& 313.335 & 293.351 & 238.027 & 183.198 & 186.125 \\
BVH~\cite{Matt:2016:Pbr}& 198.140 & 227.333 & 243.165 & 177.263 & 187.540 \\
\textit{k}-d tree~\cite{Matt:2016:Pbr}& \textbf{119.589} & \textbf{127.948} & \textbf{126.434} & \textbf{114.358} & \textbf{121.114} \\
\hline
\multicolumn{6}{c}{}\\
\end{tabular}
\label{tab:build_and_render_times_remesh}
\end{table*}

%% file: render_times_gpu/table_black.tex
\begin{table*}[htbp]
\centering
\small
\caption{The rendering times of tetrahedral mesh-based acceleration structures on the GPU}
\begin{tabular}{|lrrrrrr|}
\hline
\multicolumn{7}{|c|}{Scenes}\\
\hline
& \multicolumn{1}{c}{Torus Knots}
& \multicolumn{1}{c}{Armadillo}
& \multicolumn{1}{c}{Neptune}
& \multicolumn{1}{c}{Mix}
& \multicolumn{1}{c}{Rungholt}
& \multicolumn{1}{c|}{Exhaust Pipe}
\\
& \includegraphics[width = 0.115\textwidth]{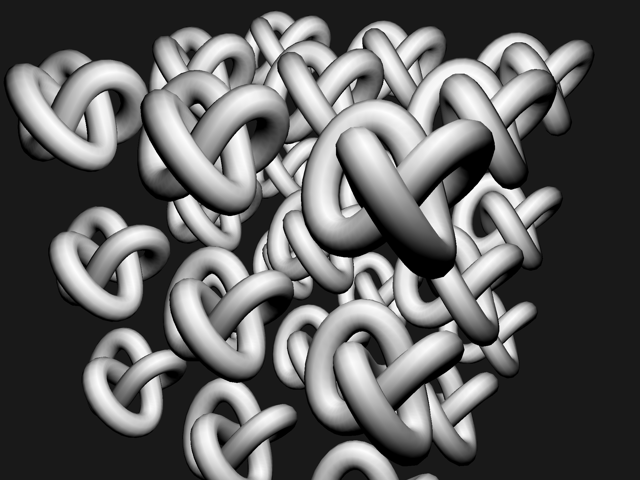} & \includegraphics[width = 0.115\textwidth]{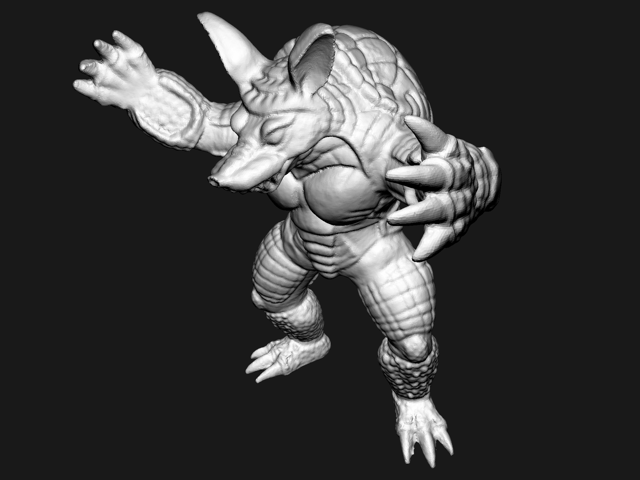} & \includegraphics[width = 0.115\textwidth]{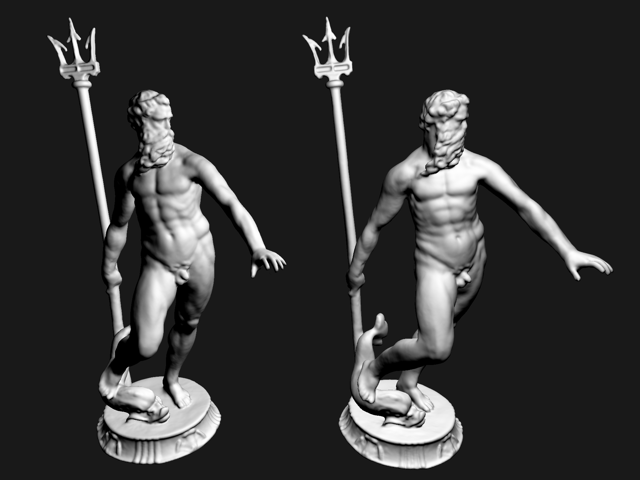} & \includegraphics[width = 0.115\textwidth]{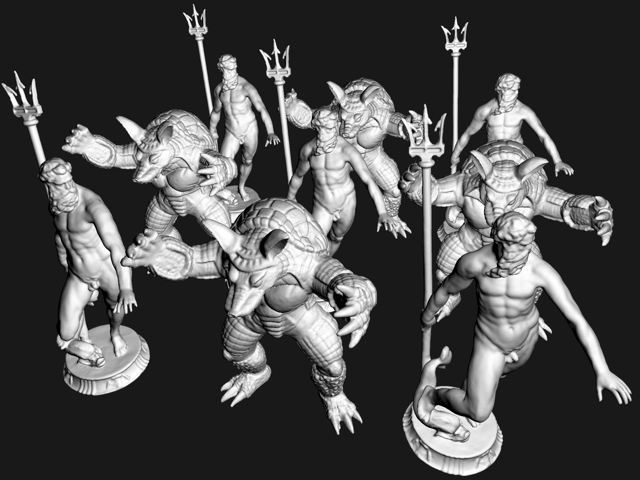} &
\includegraphics[width = 0.115\textwidth]{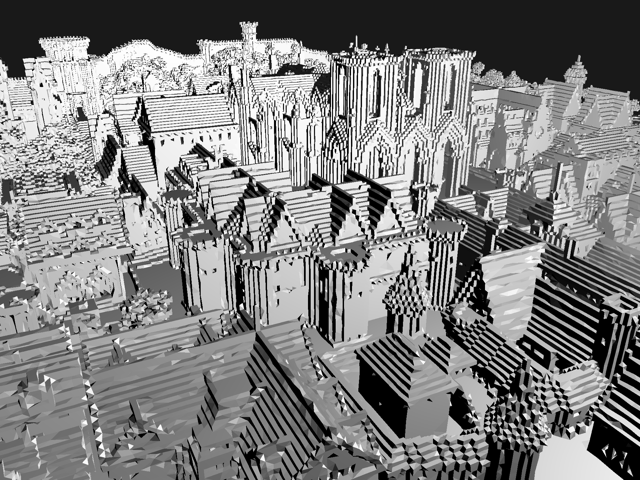} &
\includegraphics[width = 0.115\textwidth]{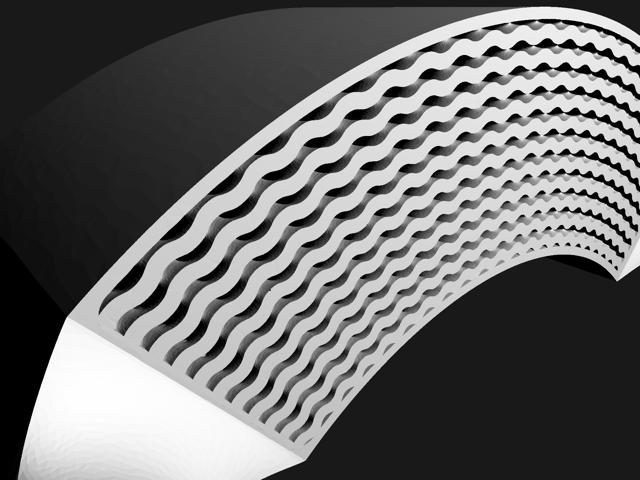} \\
\hline
\hline
\multicolumn{7}{|c|}{Scene statistics} \\
\hline
\# of triangles & 77,760 & 345,938 & 448,896 & 2,505,992 & 3,580,928 & 6,244,678 \\
\hline
\hline
\multicolumn{7}{|c|}{Kernel execution time (in milliseconds)} \\
\hline
Tet-mesh-ScTP~\cite{Lagae:2008:Constrained}& 20.021	& 19.612 & 20.898 & 43.910 & 43.633 & 22.560 \\
Tet-mesh-80~\cite{Maria:2017:Traversal}& 7.790 & 7.136 & 7.941 & 13.454 & 14.360 & 9.023 \\
Tet-mesh-32& 19.541	& 18.950 & 20.958 & 42.690 & 44.544 & 21.320 \\
Tet-mesh-20& \textbf{6.156} & \textbf{5.803} & \textbf{6.529} & \textbf{11.322} & \textbf{12.172} & \textbf{6.931} \\
Tet-mesh-16& 7.120 & 6.477 & 7.328 & 12.157 & 13.444 & 8.231 \\

\hline
\end{tabular}
\label{tab:render_times_gpu}
\end{table*}

%% file: memory_usage/table_black.tex
\begin{table*}[ht]
\centering
\small
\caption{The memory requirements of tetrahedral mesh-based acceleration structures}
\begin{tabular}{|lrrrrrr|}
\hline
\multicolumn{7}{|c|}{Scenes}\\
\hline
& \multicolumn{1}{c}{Torus Knots}
& \multicolumn{1}{c}{Armadillo}
& \multicolumn{1}{c}{Neptune}
& \multicolumn{1}{c}{Mix}
& \multicolumn{1}{c}{Rungholt}
& \multicolumn{1}{c|}{Exhaust Pipe}
\\
& \includegraphics[width = 0.12\textwidth]{{"memory_usage/Torus_Knots"}.png} & \includegraphics[width = 0.12\textwidth]{{"memory_usage/Armadillo"}.png} & \includegraphics[width = 0.12\textwidth]{{"memory_usage/Neptune"}.png} & \includegraphics[width = 0.12\textwidth]{{"memory_usage/Mix"}.png} &
\includegraphics[width = 0.12\textwidth]{{"memory_usage/Rungholt_Default"}.png} &
\includegraphics[width = 0.12\textwidth]{{"memory_usage/Exhaust_Pipe_Left"}.png} \\
\hline
\hline
\multicolumn{7}{|c|}{Scene statistics} \\
\hline
\# of triangles & 77,760 & 345,938 & 448,896 & 2,505,992 & 3,580,928 & 6,244,678\\
\hline
\hline
\multicolumn{7}{|c|}{Accelerator size (in megabytes)} \\
\hline
Tet-mesh-32& 12.3 & 49.4 & 61.2 & 352.1 & 406.1 & 885.1 \\
Tet-mesh-20& 9.2 & 37.7 & 47.0 & 269.0 & 310.2 & 673.6 \\
Tet-mesh-16& \textbf{8.2} & \textbf{33.7} & \textbf{42.2} & \textbf{241.3} & \textbf{278.2} & \textbf{603.1} \\
Tet-mesh-80~\cite{Maria:2017:Traversal}& 20.6 & 78.4 & 94.6 & 553.8 & 639.4 & 1410.0 \\
\hline
\end{tabular}
\label{tab:memory_usage}
\end{table*}

%% file: camera_distance/table.tex
\begin{multicols}{2}
\begin{table*}
\centering
\small
\caption{The rendering times and visited node counts for different types of accelerators as the camera gets closer to the mesh surface}
\begin{tabular}{|lrrrrrr|}
\hline
\multicolumn{7}{|c|} {Scenes} \\
\hline
\rule{0pt}{13.5ex} Tet-mesh-20 & \includegraphics[width = 0.12\textwidth]{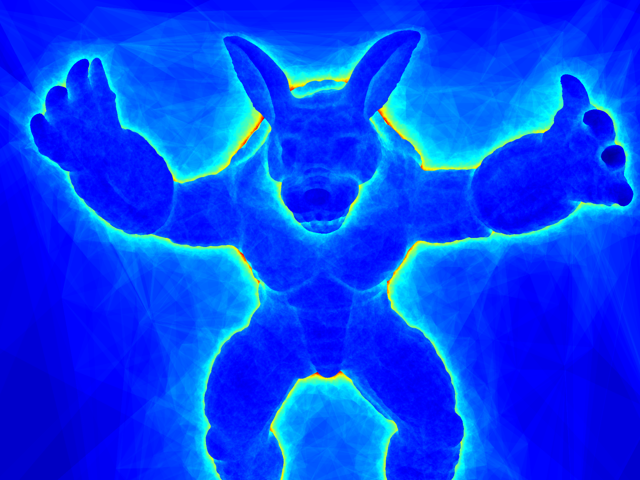} & \includegraphics[width = 0.12\textwidth]{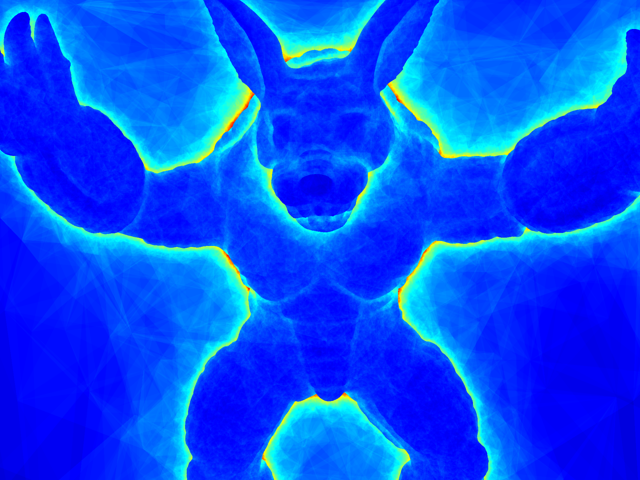} & \includegraphics[width = 0.12\textwidth]{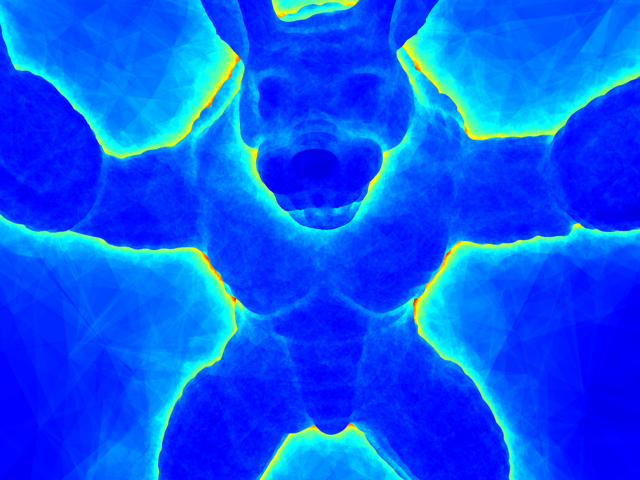} & \includegraphics[width = 0.12\textwidth]{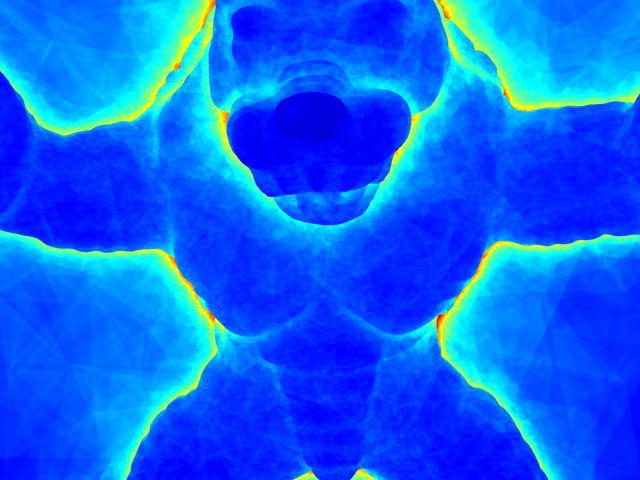} & \includegraphics[width = 0.12\textwidth]{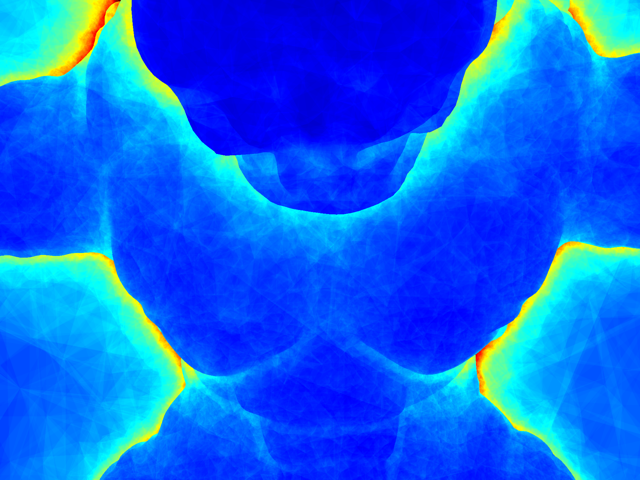} & \includegraphics[width = 0.12\textwidth]{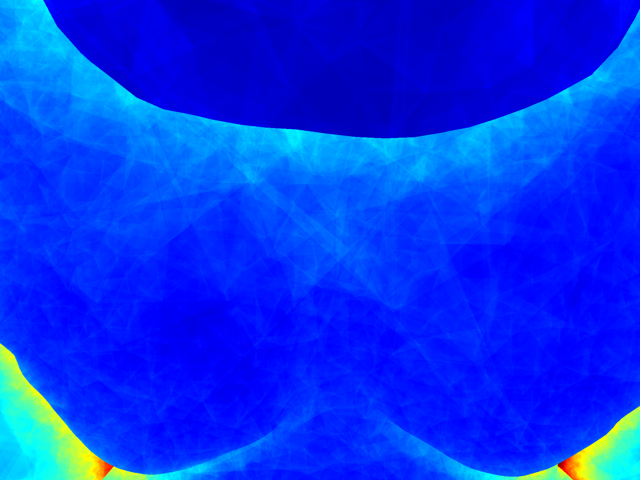} \\
\hline
\rule{0pt}{13.5ex} BVH~\cite{Matt:2016:Pbr} & \includegraphics[width = 0.12\textwidth]{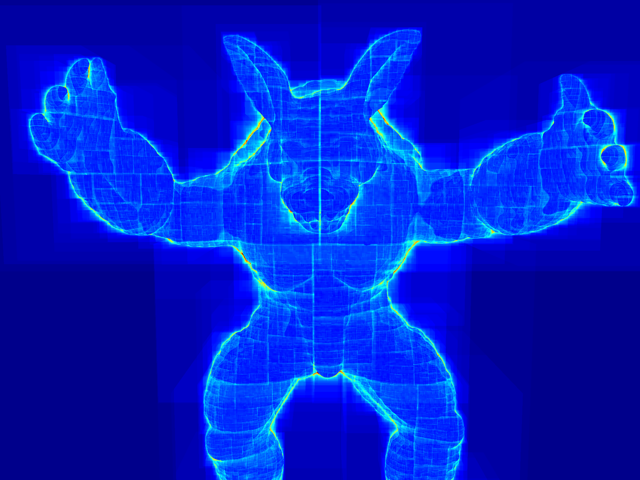} & \includegraphics[width = 0.12\textwidth]{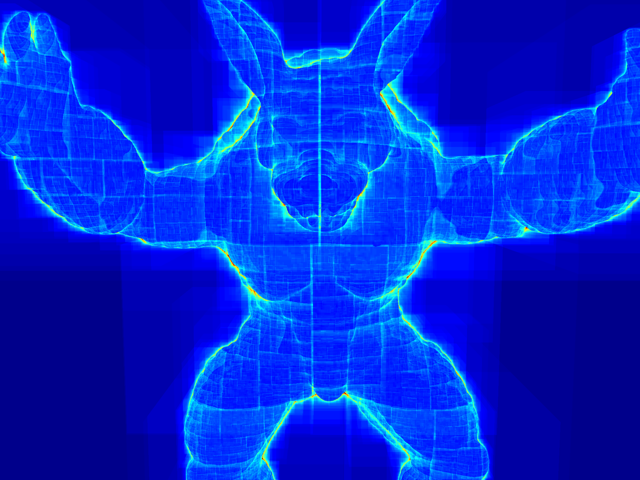} & \includegraphics[width = 0.12\textwidth]{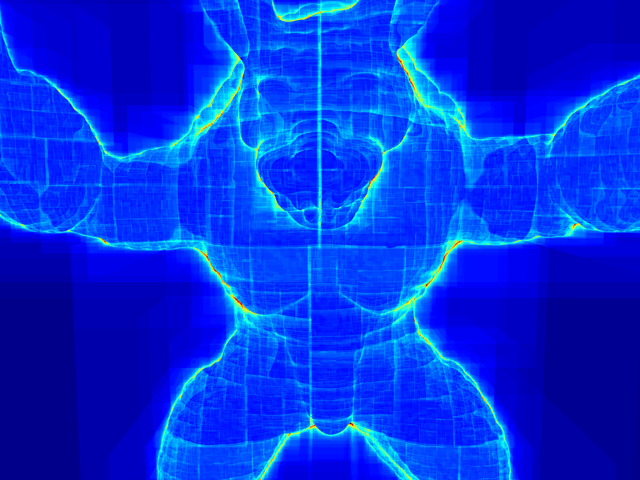} & \includegraphics[width = 0.12\textwidth]{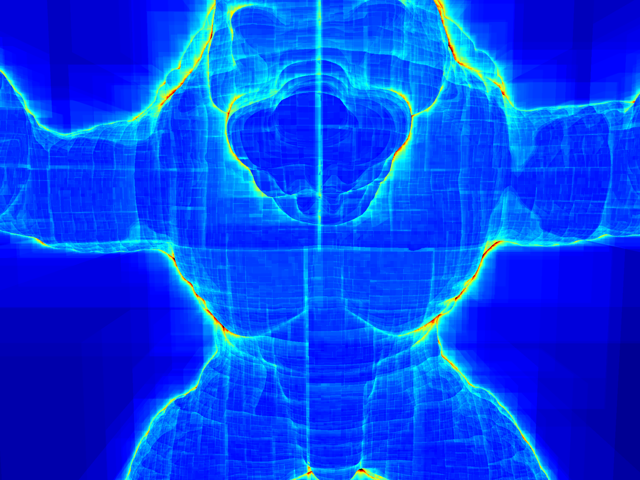} & \includegraphics[width = 0.12\textwidth]{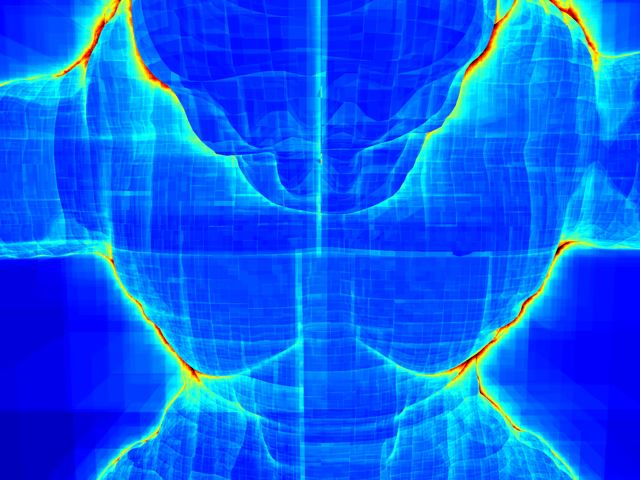} & \includegraphics[width = 0.12\textwidth]{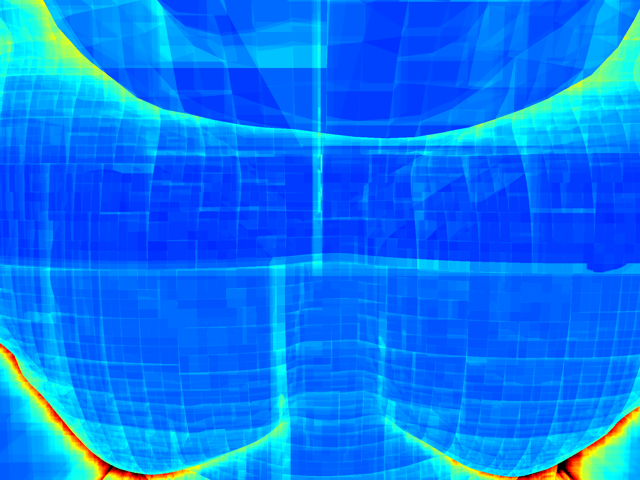}\\
\hline
\rule{0pt}{13.5ex} Kd-tree~\cite{Matt:2016:Pbr} & \includegraphics[width = 0.12\textwidth]{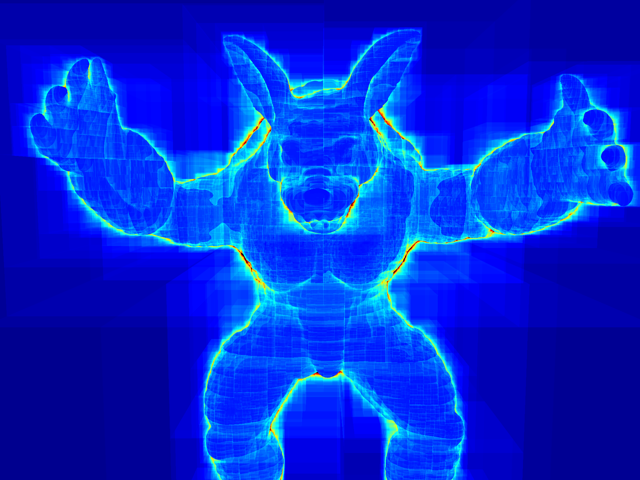} & \includegraphics[width = 0.12\textwidth]{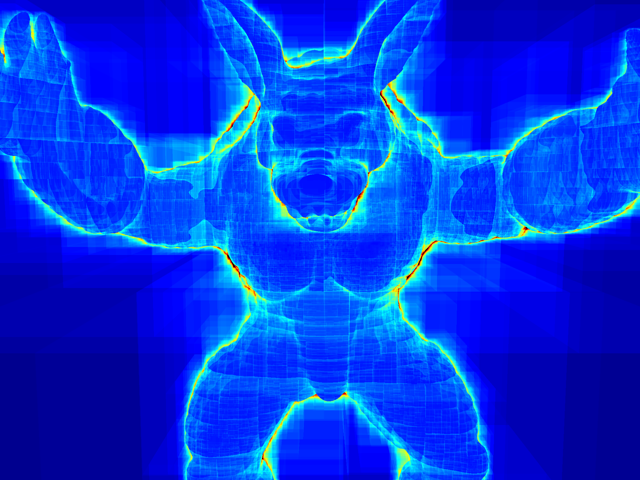} & \includegraphics[width = 0.12\textwidth]{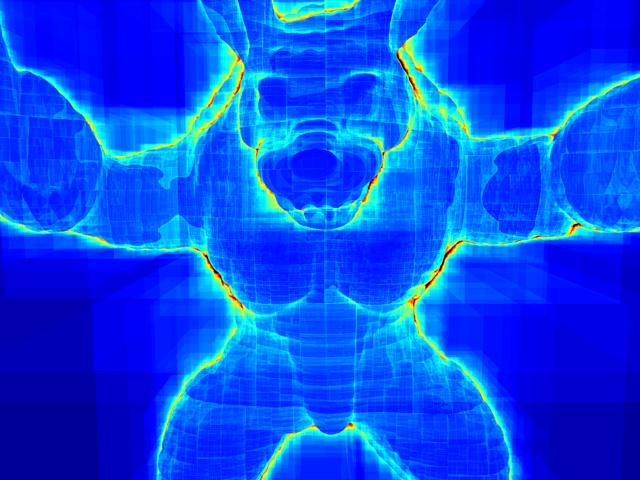} & \includegraphics[width = 0.12\textwidth]{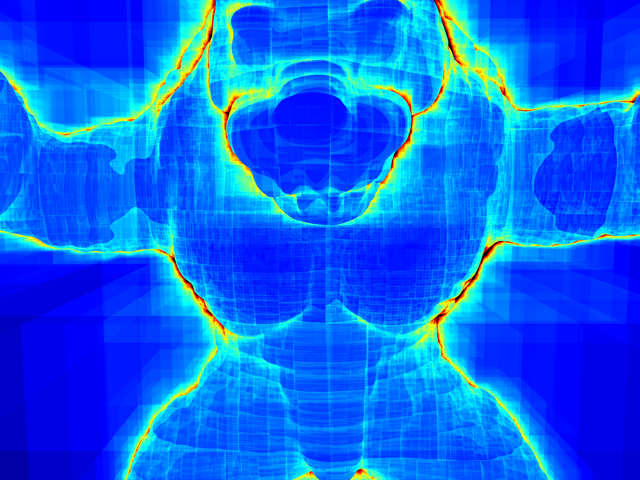} & \includegraphics[width = 0.12\textwidth]{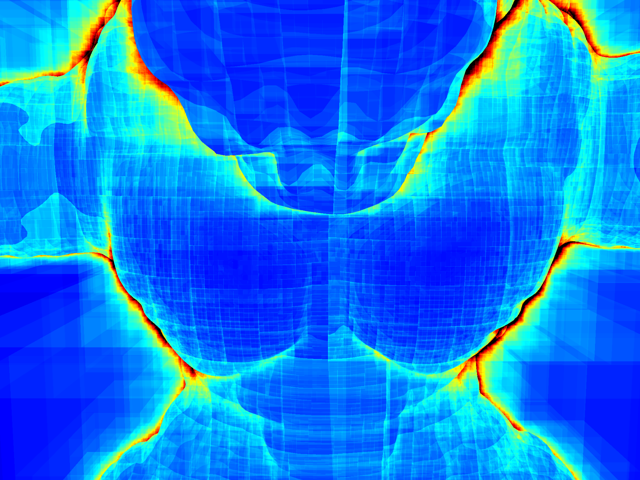} & \includegraphics[width = 0.12\textwidth]{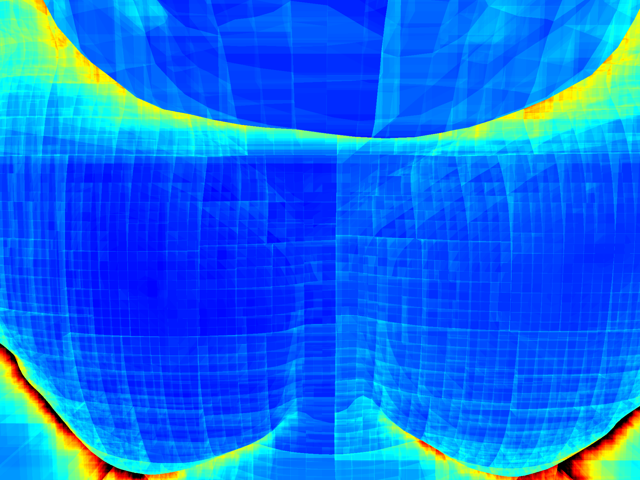}\\
\hline
\hline
\multicolumn{7}{|c|} {Visited node count per pixel} \\
\hline
Tet-mesh-20& 48.54 & 52.32 & 55.11 & 59.13 & 60.13 & \textbf{43.62} \\
BVH~\cite{Matt:2016:Pbr}& \textbf{27.23} & \textbf{32.90} & \textbf{38.53} & \textbf{46.67} & \textbf{57.88} & 65.27 \\
\textit{k}-d tree~\cite{Matt:2016:Pbr}& 34.12 & 41.84 & 49.50 & 60.21 & 70.18 & 65.46 \\
\hline
\hline
\multicolumn{7}{|c|} {Rendering times (in milliseconds)} \\
\hline
Tet-mesh-20& 140.3 & 151.0 & 168.3 & 179.4 & 182.3 & \textbf{126.6} \\
BVH~\cite{Matt:2016:Pbr}& 87.4 & 109.4 & 123.9 & 146.8 & 175.9 & 193.9 \\
\textit{k}-d tree~\cite{Matt:2016:Pbr}& \textbf{86.8} & \textbf{107.4} & \textbf{118.8} & \textbf{136.8} & \textbf{157.9} & 144.0 \\
\hline
\end{tabular}
\label{tab:camera_distance}
\end{table*}
\end{multicols}

%% file: arxiv_tech_report.bbl
\begin{thebibliography}{40}
\providecommand{\natexlab}[1]{#1}
\providecommand{\url}[1]{\texttt{#1}}
\providecommand{\href}[2]{#2}
\providecommand{\path}[1]{#1}
\providecommand{\eprint}[1]{\href{http://arxiv.org/abs/#1}{\path{#1}}}
\providecommand{\DOIprefix}{doi:}
\providecommand{\ArXivprefix}{arXiv:}
\providecommand{\URLprefix}{URL: }
\providecommand{\Pubmedprefix}{pmid:}
\providecommand{\doi}[1]{\href{http://dx.doi.org/#1}{\path{#1}}}
\providecommand{\Pubmed}[1]{\href{pmid:#1}{\path{#1}}}
\providecommand{\BIBand}{and}
\providecommand{\bibinfo}[2]{#2}
\ifx\xfnm\undefined \def\xfnm[#1]{\unskip,\space#1}\fi
\bibitem[{Glassner(1989)}]{Glassner:1989:Raytracing}
\bibinfo{editor}{Glassner\xfnm[ AS]}, editor.
\newblock \bibinfo{title}{{An Introduction to Ray Tracing}}.
\newblock \bibinfo{address}{London, UK}: \bibinfo{publisher}{Academic Press
  Ltd.}; \bibinfo{year}{1989}.
\bibitem[{Lagae and Dutr\'{e}(2008{\natexlab{a}})}]{Lagae:2008:Constrained}
\bibinfo{author}{Lagae\xfnm[ A]}, \bibinfo{author}{Dutr\'{e}\xfnm[ P]}.
\newblock \bibinfo{title}{Accelerating ray tracing using constrained
  tetrahedralizations}.
\newblock \bibinfo{journal}{Computer Graphics Forum}
  \bibinfo{year}{2008}{\natexlab{a}};\bibinfo{volume}{27}(\bibinfo{number}{4}):\bibinfo{pages}{1303--1312}.
\bibitem[{Pharr et~al.(2016)Pharr, Jakob and Humphreys}]{Matt:2016:Pbr}
\bibinfo{author}{Pharr\xfnm[ M]}, \bibinfo{author}{Jakob\xfnm[ W]},
  \bibinfo{author}{Humphreys\xfnm[ G]}.
\newblock \bibinfo{title}{Physically Based Rendering: From Theory to
  Implementation}.
\newblock \bibinfo{edition}{3rd} ed.; \bibinfo{address}{San Francisco, CA,
  USA}: \bibinfo{publisher}{Morgan Kaufmann Publishers, Inc.};
  \bibinfo{year}{2016}.
\bibitem[{Fujimoto et~al.(1988)Fujimoto, Tanaka and
  Iwata}]{Fujimoto:1988:Accelerated}
\bibinfo{author}{Fujimoto\xfnm[ A]}, \bibinfo{author}{Tanaka\xfnm[ T]},
  \bibinfo{author}{Iwata\xfnm[ K]}.
\newblock \bibinfo{title}{{ARTS: Accelerated Ray-tracing System}}.
\newblock In: \bibinfo{editor}{Joy\xfnm[ KI]}, \bibinfo{editor}{Grant\xfnm[
  CW]}, \bibinfo{editor}{Max\xfnm[ NL]}, \bibinfo{editor}{Hatfield\xfnm[ L]},
  editors. \bibinfo{booktitle}{{Tutorial: Computer Graphics; Image Synthesis}}.
  \bibinfo{address}{New York, NY, USA}: \bibinfo{publisher}{Computer Science
  Press, Inc.}; \bibinfo{year}{1988}, p. \bibinfo{pages}{148--159}.
\bibitem[{Lagae and Dutr\'{e}(2008{\natexlab{b}})}]{Lagae:2008:Grid}
\bibinfo{author}{Lagae\xfnm[ A]}, \bibinfo{author}{Dutr\'{e}\xfnm[ P]}.
\newblock \bibinfo{title}{Compact, fast and robust grids for ray tracing}.
\newblock \bibinfo{journal}{Computer Graphics Forum}
  \bibinfo{year}{2008}{\natexlab{b}};\bibinfo{volume}{27}(\bibinfo{number}{4}):\bibinfo{pages}{1235--1244}.
\bibitem[{Goldsmith and Salmon(1987)}]{Goldsmith:1987:Hierarchies}
\bibinfo{author}{Goldsmith\xfnm[ J]}, \bibinfo{author}{Salmon\xfnm[ J]}.
\newblock \bibinfo{title}{{Automatic Creation of Object Hierarchies for Ray
  Tracing}}.
\newblock \bibinfo{journal}{IEEE Computer Graphics and Applications}
  \bibinfo{year}{1987};\bibinfo{volume}{7}(\bibinfo{number}{5}):\bibinfo{pages}{14--20}.
\bibitem[{MacDonald and Booth(1990)}]{MacDonald:1990:Subdivision}
\bibinfo{author}{MacDonald\xfnm[ DJ]}, \bibinfo{author}{Booth\xfnm[ KS]}.
\newblock \bibinfo{title}{{Heuristics for Ray Tracing Using Space
  Subdivision}}.
\newblock \bibinfo{journal}{The Visual Computer}
  \bibinfo{year}{1990};\bibinfo{volume}{6}(\bibinfo{number}{3}):\bibinfo{pages}{153--166}.
\bibitem[{Stich et~al.(2009)Stich, Friedrich and Dietrich}]{Stich:2009:Splits}
\bibinfo{author}{Stich\xfnm[ M]}, \bibinfo{author}{Friedrich\xfnm[ H]},
  \bibinfo{author}{Dietrich\xfnm[ A]}.
\newblock \bibinfo{title}{{Spatial Splits in Bounding Volume Hierarchies}}.
\newblock In: \bibinfo{booktitle}{Proceedings of the Conference on High
  Performance Graphics}. HPG~'09; \bibinfo{address}{New York, NY, USA}:
  \bibinfo{publisher}{ACM}; \bibinfo{year}{2009}, p. \bibinfo{pages}{7--13}.
\bibitem[{Wodniok and Goesele(2017)}]{Wodniok:2016:Subtree}
\bibinfo{author}{Wodniok\xfnm[ D]}, \bibinfo{author}{Goesele\xfnm[ M]}.
\newblock \bibinfo{title}{{Construction of Bounding Volume Hierarchies with
  {SAH} Cost Approximation on Temporary Subtrees}}.
\newblock \bibinfo{journal}{Computers \& Graphics}
  \bibinfo{year}{2017};\bibinfo{volume}{62}:\bibinfo{pages}{41--52}.
\bibitem[{Glassner(1984)}]{Glassner:1984:Subdivision}
\bibinfo{author}{Glassner\xfnm[ AS]}.
\newblock \bibinfo{title}{Space subdivision for fast ray tracing}.
\newblock \bibinfo{journal}{IEEE Computer Graphics and Applications}
  \bibinfo{year}{1984};\bibinfo{volume}{4}(\bibinfo{number}{10}):\bibinfo{pages}{15--24}.
\bibitem[{Havran and Bittner(2002)}]{Havran:2002:RayShooting}
\bibinfo{author}{Havran\xfnm[ V]}, \bibinfo{author}{Bittner\xfnm[ J]}.
\newblock \bibinfo{title}{On improving kd tree for ray shooting}.
\newblock \bibinfo{journal}{Journal of WSCG}
  \bibinfo{year}{2002};\bibinfo{volume}{10}:\bibinfo{pages}{209--216}.
\bibitem[{Wald and Havran(2006)}]{Wald:2006:FastKd}
\bibinfo{author}{Wald\xfnm[ I]}, \bibinfo{author}{Havran\xfnm[ V]}.
\newblock \bibinfo{title}{{On Building Fast kd-Trees for Ray Tracing, and on
  Doing That in O(N log N)}}.
\newblock In: \bibinfo{booktitle}{Proceedings of the IEEE Symposium on
  Interactive Ray Tracing}. \bibinfo{year}{2006}, p. \bibinfo{pages}{61--69}.
\bibitem[{Haines and Greenberg(1986)}]{Haines:1986:LightBuffer}
\bibinfo{author}{Haines\xfnm[ E]}, \bibinfo{author}{Greenberg\xfnm[ D]}.
\newblock \bibinfo{title}{{The Light Buffer: A Shadow-Testing Accelerator}}.
\newblock \bibinfo{journal}{IEEE Computer Graphics and Applications}
  \bibinfo{year}{1986};\bibinfo{volume}{6}(\bibinfo{number}{9}):\bibinfo{pages}{6--16}.
\bibitem[{Hunt and Mark(2008{\natexlab{a}})}]{Hunt:2008:Adaptive}
\bibinfo{author}{Hunt\xfnm[ W]}, \bibinfo{author}{Mark\xfnm[ W]}.
\newblock \bibinfo{title}{{Adaptive Acceleration Structures in Perspective
  Space}}.
\newblock In: \bibinfo{booktitle}{Proceedings of the IEEE Symposium on
  Interactive Ray Tracing}. RT~'08; \bibinfo{year}{2008}{\natexlab{a}}, p.
  \bibinfo{pages}{11--17}.
\bibitem[{Hunt and Mark(2008{\natexlab{b}})}]{Hunt:2008:RaySpecialized}
\bibinfo{author}{Hunt\xfnm[ W]}, \bibinfo{author}{Mark\xfnm[ W]}.
\newblock \bibinfo{title}{{Ray-specialized Acceleration Structures for Ray
  Tracing}}.
\newblock In: \bibinfo{booktitle}{Proceedings of the IEEE Symposium on
  Interactive Ray Tracing}. RT~'08; \bibinfo{year}{2008}{\natexlab{b}}, p.
  \bibinfo{pages}{3--10}.
\bibitem[{Si(2015)}]{Si:2015:TetGen}
\bibinfo{author}{Si\xfnm[ H]}.
\newblock \bibinfo{title}{{TetGen, a Delaunay}-based quality tetrahedral mesh
  generator}.
\newblock \bibinfo{journal}{ACM Transactions on Mathematical Software}
  \bibinfo{year}{2015};\bibinfo{volume}{41}(\bibinfo{number}{2}):\bibinfo{pages}{11:1--11:36}.
\bibitem[{Bowyer(1981)}]{Bowyer:1981:Dirichlet}
\bibinfo{author}{Bowyer\xfnm[ A]}.
\newblock \bibinfo{title}{Computing {Dirichlet} tessellations}.
\newblock \bibinfo{journal}{The Computer Journal}
  \bibinfo{year}{1981};\bibinfo{volume}{24}(\bibinfo{number}{2}):\bibinfo{pages}{162--166}.
\bibitem[{Watson(1981)}]{Watson:1981:Delaunay}
\bibinfo{author}{Watson\xfnm[ DF]}.
\newblock \bibinfo{title}{Computing the n-dimensional {Delaunay} tessellation
  with application to {Voronoi} polytopes}.
\newblock \bibinfo{journal}{The Computer Journal}
  \bibinfo{year}{1981};\bibinfo{volume}{24}(\bibinfo{number}{2}):\bibinfo{pages}{167--172}.
\bibitem[{Edelsbrunner and Shah(1992)}]{Edelsbrunner:1992:Flipping}
\bibinfo{author}{Edelsbrunner\xfnm[ H]}, \bibinfo{author}{Shah\xfnm[ NR]}.
\newblock \bibinfo{title}{{Incremental Topological Flipping Works for Regular
  Triangulations}}.
\newblock In: \bibinfo{booktitle}{Proceedings of the Eighth Annual Symposium on
  Computational Geometry}. SCG~'92; \bibinfo{address}{New York, NY, USA}:
  \bibinfo{publisher}{ACM}; \bibinfo{year}{1992}, p. \bibinfo{pages}{43--52}.
\bibitem[{Shewchuk(1996)}]{Shewchuk:1996:Predicates}
\bibinfo{author}{Shewchuk\xfnm[ JR]}.
\newblock \bibinfo{title}{{Adaptive Precision Floating-Point Arithmetic and
  Fast Robust Geometric Predicates}}.
\newblock \bibinfo{journal}{Discrete and Computational Geometry}
  \bibinfo{year}{1996};\bibinfo{volume}{18}:\bibinfo{pages}{305--363}.
\bibitem[{{Maria} et~al.(2017{\natexlab{a}}){Maria}, {Horna} and
  {Aveneau}}]{Maria:2017:Traversal}
\bibinfo{author}{{Maria}\xfnm[ M]}, \bibinfo{author}{{Horna}\xfnm[ S]},
  \bibinfo{author}{{Aveneau}\xfnm[ L]}.
\newblock \bibinfo{title}{Efficient ray traversal of constrained {Delaunay}
  tetrahedralization}.
\newblock In: \bibinfo{booktitle}{Proceedings of the 12th International Joint
  Conference on Computer Vision, Imaging and Computer Graphics Theory and
  Applications}; vol.~\bibinfo{volume}{1} of
  \emph{\bibinfo{series}{VISIGRAPP~'17}}. \bibinfo{year}{2017}{\natexlab{a}},
  p. \bibinfo{pages}{236--243}.
\bibitem[{{Maria} et~al.(2017{\natexlab{b}}){Maria}, {Horna} and
  {Aveneau}}]{Maria:2017:Convex}
\bibinfo{author}{{Maria}\xfnm[ M]}, \bibinfo{author}{{Horna}\xfnm[ S]},
  \bibinfo{author}{{Aveneau}\xfnm[ L]}.
\newblock \bibinfo{title}{Constrained convex space partition for ray tracing in
  architectural environments}.
\newblock \bibinfo{journal}{Computer Graphics Forum}
  \bibinfo{year}{2017}{\natexlab{b}};\bibinfo{volume}{36}(\bibinfo{number}{1}):\bibinfo{pages}{288--300}.
\bibitem[{{Maria} et~al.(2014){Maria}, {Horna} and
  {Aveneau}}]{Maria:2014:Topological}
\bibinfo{author}{{Maria}\xfnm[ M]}, \bibinfo{author}{{Horna}\xfnm[ S]},
  \bibinfo{author}{{Aveneau}\xfnm[ L]}.
\newblock \bibinfo{title}{Topological space partition for fast ray tracing in
  architectural models}.
\newblock In: \bibinfo{booktitle}{Proceedings of the International Conference
  on Computer Graphics Theory and Applications}. GRAPP~'14;
  \bibinfo{year}{2014}, p. \bibinfo{pages}{1--11}.
\bibitem[{{Silva} et~al.(1996){Silva}, {Mitchell} and
  {Kaufman}}]{Silva:1996:LazySweep}
\bibinfo{author}{{Silva}\xfnm[ CT]}, \bibinfo{author}{{Mitchell}\xfnm[ JSB]},
  \bibinfo{author}{{Kaufman}\xfnm[ AE]}.
\newblock \bibinfo{title}{{Fast Rendering of Irregular Grids}}.
\newblock In: \bibinfo{booktitle}{Proceedings of Symposium on Volume
  Visualization}. \bibinfo{year}{1996}, p. \bibinfo{pages}{15--22}.
\bibitem[{{Silva} and {Mitchell}(1997)}]{Silva:1997:LazySweep}
\bibinfo{author}{{Silva}\xfnm[ CT]}, \bibinfo{author}{{Mitchell}\xfnm[ JSB]}.
\newblock \bibinfo{title}{{The Lazy Sweep Ray Casting Algorithm for Rendering
  Irregular Grids}}.
\newblock \bibinfo{journal}{IEEE Transactions on Visualization and Computer
  Graphics}
  \bibinfo{year}{1997};\bibinfo{volume}{3}(\bibinfo{number}{2}):\bibinfo{pages}{142--157}.
\bibitem[{Berk et~al.(2003)Berk, Aykanat and Gudukbay}]{Berk:2003:Direct}
\bibinfo{author}{Berk\xfnm[ H]}, \bibinfo{author}{Aykanat\xfnm[ C]},
  \bibinfo{author}{Gudukbay\xfnm[ U]}.
\newblock \bibinfo{title}{Direct volume rendering of unstructured grids}.
\newblock \bibinfo{journal}{Computers \& Graphics}
  \bibinfo{year}{2003};\bibinfo{volume}{27}:\bibinfo{pages}{387--406}.
\bibitem[{Koyamada(1992)}]{Koyamada:1992:Fast}
\bibinfo{author}{Koyamada\xfnm[ K]}.
\newblock \bibinfo{title}{Fast traverse of irregular volumes}.
\newblock In: \bibinfo{editor}{Kunii\xfnm[ TL]}, editor.
  \bibinfo{booktitle}{Visual Computing}. \bibinfo{address}{Tokyo}:
  \bibinfo{publisher}{Springer Japan}; \bibinfo{year}{1992}, p.
  \bibinfo{pages}{295--311}.
\bibitem[{Garrity(1990)}]{Garrity:1990:Irregular}
\bibinfo{author}{Garrity\xfnm[ MP]}.
\newblock \bibinfo{title}{Raytracing irregular volume data}.
\newblock In: \bibinfo{booktitle}{Proceedings of the Workshop on Volume
  Visualization}. VVS~'90; \bibinfo{address}{New York, NY, USA}:
  \bibinfo{publisher}{ACM}; \bibinfo{year}{1990}, p. \bibinfo{pages}{35--40}.
\bibitem[{{Ribeiro} et~al.(2007){Ribeiro}, {Maximo}, {Bentes}, {Oliveira} and
  {Farias}}]{Ribeiro:2007:Memory}
\bibinfo{author}{{Ribeiro}\xfnm[ S]}, \bibinfo{author}{{Maximo}\xfnm[ A]},
  \bibinfo{author}{{Bentes}\xfnm[ C]}, \bibinfo{author}{{Oliveira}\xfnm[ A]},
  \bibinfo{author}{{Farias}\xfnm[ R]}.
\newblock \bibinfo{title}{Memory-aware and efficient ray-casting algorithm}.
\newblock In: \bibinfo{booktitle}{Proceedings of the XX Brazilian Symposium on
  Computer Graphics and Image Processing}. SIBGRAPI~'07; \bibinfo{year}{2007},
  p. \bibinfo{pages}{147--154}.
\bibitem[{Maximo et~al.(2008)Maximo, Ribeiro, Bentes, Oliveira and
  Farias}]{Maximo:2008:Memory}
\bibinfo{author}{Maximo\xfnm[ A]}, \bibinfo{author}{Ribeiro\xfnm[ S]},
  \bibinfo{author}{Bentes\xfnm[ C]}, \bibinfo{author}{Oliveira\xfnm[ A]},
  \bibinfo{author}{Farias\xfnm[ R]}.
\newblock \bibinfo{title}{Memory efficient {gpu}-based ray casting for
  unstructured volume rendering}.
\newblock In: \bibinfo{booktitle}{Proceedings of the Fifth Eurographics / IEEE
  VGTC Conference on Point-Based Graphics}. SPBG’08;
  \bibinfo{address}{Goslar, DEU}: \bibinfo{publisher}{Eurographics
  Association}; \bibinfo{year}{2008}, p. \bibinfo{pages}{155--162}.
\bibitem[{Marmitt and Slusallek(2006)}]{Marmitt:2006:Traversal}
\bibinfo{author}{Marmitt\xfnm[ G]}, \bibinfo{author}{Slusallek\xfnm[ P]}.
\newblock \bibinfo{title}{Fast ray traversal of tetrahedral and hexahedral
  meshes for direct volume rendering}.
\newblock In: \bibinfo{booktitle}{Proceedings of the Eighth Joint Eurographics
  / IEEE VGTC Conference on Visualization}. EUROVIS~'06;
  \bibinfo{address}{Aire-la-Ville, Switzerland}:
  \bibinfo{publisher}{Eurographics Association}; \bibinfo{year}{2006}, p.
  \bibinfo{pages}{235--242}.
\bibitem[{Platis and Theoharis(2003)}]{Platis:2003:Plucker}
\bibinfo{author}{Platis\xfnm[ N]}, \bibinfo{author}{Theoharis\xfnm[ T]}.
\newblock \bibinfo{title}{{Fast Ray-Tetrahedron Intersection Using Pl\"{u}cker
  Coordinates}}.
\newblock \bibinfo{journal}{Journal of Graphics Tools}
  \bibinfo{year}{2003};\bibinfo{volume}{8}(\bibinfo{number}{4}):\bibinfo{pages}{37--48}.
\bibitem[{Sinha(2005)}]{Sinha:2005:DoublyLinkedList}
\bibinfo{author}{Sinha\xfnm[ P]}.
\newblock \bibinfo{title}{A memory-efficient doubly linked list}.
\newblock \bibinfo{journal}{Linux Journal}
  \bibinfo{year}{2005};\bibinfo{volume}{2005}(\bibinfo{number}{129}):\bibinfo{pages}{10}.
\bibitem[{Mebarki(2018)}]{Mebarki:2018:Xor}
\bibinfo{author}{Mebarki\xfnm[ A]}.
\newblock \bibinfo{title}{{XOR}-based compact triangulations}.
\newblock \bibinfo{journal}{Computing and Informatics}
  \bibinfo{year}{2018};\bibinfo{volume}{37}:\bibinfo{pages}{367--384}.
\bibitem[{Duff et~al.(2017)Duff, Burgess, Christensen, Hery, Kensler, Liani
  et~al.}]{Duff:2017:Orthonormal}
\bibinfo{author}{Duff\xfnm[ T]}, \bibinfo{author}{Burgess\xfnm[ J]},
  \bibinfo{author}{Christensen\xfnm[ P]}, \bibinfo{author}{Hery\xfnm[ C]},
  \bibinfo{author}{Kensler\xfnm[ A]}, \bibinfo{author}{Liani\xfnm[ M]}, et~al.
\newblock \bibinfo{title}{Building an orthonormal basis, revisited}.
\newblock \bibinfo{journal}{Journal of Computer Graphics Techniques}
  \bibinfo{year}{2017};\bibinfo{volume}{6}(\bibinfo{number}{1}):\bibinfo{pages}{1--8}.
\bibitem[{Wald(2007)}]{Wald:2007:FastSAH}
\bibinfo{author}{Wald\xfnm[ I]}.
\newblock \bibinfo{title}{On fast construction of {SAH}-based bounding volume
  hierarchies}.
\newblock In: \bibinfo{booktitle}{Proceedings of the IEEE Symposium on
  Interactive Ray Tracing}. RT~'07; \bibinfo{address}{Washington, DC, USA}:
  \bibinfo{publisher}{IEEE Computer Society}; \bibinfo{year}{2007}, p.
  \bibinfo{pages}{33--40}.


  
\bibitem[{Gunther et~al.(2007)Gunther, Popov, Seidel and
  Slusallek}]{Gunther:2007:Packet}
\bibinfo{author}{Gunther\xfnm[ J]}, \bibinfo{author}{Popov\xfnm[ S]},
  \bibinfo{author}{Seidel\xfnm[ HP]}, \bibinfo{author}{Slusallek\xfnm[ P]}.
\newblock \bibinfo{title}{Realtime ray tracing on {GPU with BVH}-based packet
  traversal}.
\newblock In: \bibinfo{booktitle}{Proceedings of the IEEE Symposium on
  Interactive Ray Tracing}. RT~'07; \bibinfo{address}{Washington, DC, USA}:
  \bibinfo{publisher}{IEEE Computer Society}; \bibinfo{year}{2007}, p.
  \bibinfo{pages}{113--118}.

\bibitem[{Hu et~al.(2018)Hu, Zhou, Gao, Jacobson, Zorin and
  Panozzo}]{Hu:2018:TetWild}
\bibinfo{author}{Hu\xfnm[ Y]}, \bibinfo{author}{Zhou\xfnm[ Q]},
  \bibinfo{author}{Gao\xfnm[ X]}, \bibinfo{author}{Jacobson\xfnm[ A]},
  \bibinfo{author}{Zorin\xfnm[ D]}, \bibinfo{author}{Panozzo\xfnm[ D]}.
\newblock \bibinfo{title}{Tetrahedral meshing in the wild}.
\newblock \bibinfo{journal}{ACM Transactions on Graphics}
  \bibinfo{year}{2018};\bibinfo{volume}{37}(\bibinfo{number}{4}):\bibinfo{pages}{60:1--60:14}.

  
\bibitem[{Miller et~al.(1996)Miller, Talmor, Teng, Walkington and
  Wang}]{Miller:1998:VolumeMeshes}
\bibinfo{author}{Miller\xfnm[ GL]}, \bibinfo{author}{Talmor\xfnm[ D]},
  \bibinfo{author}{Teng\xfnm[ SH]}, \bibinfo{author}{Walkington\xfnm[ N]},
  \bibinfo{author}{Wang\xfnm[ H]}.
\newblock \bibinfo{title}{Control volume meshes using sphere packing:
  Generation, refinement and coarsening}.
\newblock In: \bibinfo{booktitle}{Proceedings of the 5th International Meshing
  Roundtable}. \bibinfo{year}{1996}, p. \bibinfo{pages}{47--61}.
  
\bibitem[{Hu et~al.(2019)Hu, Schneider, Gao, Zhou, Jacobson, Zorin
  et~al.}]{Hu:2019:TriWild}
\bibinfo{author}{Hu\xfnm[ Y]}, \bibinfo{author}{Schneider\xfnm[ T]},
  \bibinfo{author}{Gao\xfnm[ X]}, \bibinfo{author}{Zhou\xfnm[ Q]},
  \bibinfo{author}{Jacobson\xfnm[ A]}, \bibinfo{author}{Zorin\xfnm[ D]}, et~al.
\newblock \bibinfo{title}{{TriWild}: Robust triangulation with curve
  constraints}.
\newblock \bibinfo{journal}{ACM Transactions on Graphics}
  \bibinfo{year}{2019};\bibinfo{volume}{38}(\bibinfo{number}{4}):\bibinfo{pages}{52:1--52:15}.

\end{thebibliography}
